\documentclass[12pt,preprint]{emulateapj}

\slugcomment{Version date: \today}

\shorttitle{DXRBS Counts, Evolution, and Luminosity Functions}
\shortauthors{Padovani et al.}

\begin{document}

\title{The Deep X-Ray Radio Blazar Survey (DXRBS). III. \\ Radio Number
Counts, Evolutionary Properties, and Luminosity Function of Blazars}

\author{Paolo Padovani}
\affil{European Southern Observatory, Karl-Schwarzschild-Str. 2,
D-85748 Garching bei M\"unchen, Germany}
\email{Paolo.Padovani@eso.org}

\author{Paolo Giommi}
\affil{ASI Science Data Center, ASDC, 
c/o ESRIN, Via G. Galilei, I-00044 Frascati, Italy}

\author{Hermine Landt}
\affil{Harvard-Smithsonian CfA, 60 Garden Street, Cambridge, MA 02138, USA}

\and

\author{Eric S. Perlman}
\affil{Joint Center for Astrophysics, University of Maryland, 1000 
Hilltop Circle, Baltimore, MD 21250, USA}
\affil{Physics and Space Sciences Department, 
Florida Institute of Technology, 150 West University Boulevard	, Melbourne, 
FL 32901, USA\altaffilmark{1}}
\altaffiltext{1}{Current address}

\begin{abstract}
Our knowledge of the blazar surface densities and luminosity functions,
which are fundamental parameters, relies still on samples at relatively
high flux limits. As a result, our understanding of this rare class of
active galactic nuclei is mostly based on relatively bright and
intrinsically luminous sources. We present the radio number counts,
evolutionary properties, and luminosity functions of the faintest blazar
sample with basically complete ($\sim 95\%$) identifications. Based on the
Deep X-ray Radio Blazar Survey (DXRBS), it includes 129 flat-spectrum radio
quasars (FSRQ) and 24 BL Lacs down to a 5 GHz flux and power $\sim 50$ mJy
and $\sim 10^{24}$ W/Hz, respectively, an order of magnitude improvement as
compared to previously published (radio-selected) blazar samples. DXRBS
FSRQ are seen to evolve strongly, up to redshift $\approx 1.5$, above which
high-power sources show a decline in their comoving space density.  DXRBS
BL Lacs, on the other hand, do not evolve. High-energy (HBL) and low-energy
(LBL) peaked BL Lacs share the same lack of cosmological evolution, which
is at variance with some previous results. The observed luminosity
functions are in good agreement with the predictions of unified schemes,
with FSRQ getting close to their expected minimum power. Despite the fact
that the large majority of our blazars are FSRQ, BL Lacs are intrinsically
$\sim 50$ times more numerous. Finally, the relative numbers of HBL and LBL
in the radio and X-ray bands are different from those predicted
by the so-called "blazar sequence" and support a scenario in which HBL
represent a small minority ($\approx 10\%$) of all BL Lacs.
\end{abstract}


\section{Introduction}

Blazars are one of the most extreme classes of active galactic nuclei
(AGN), distinguished by high luminosity, rapid variability, high
polarization, radio core-dominance (and therefore flat [$\alpha_{\rm r} \la
0.5$] radio spectra), and apparent superluminal speeds. Their broad-band
emission extends from the radio up to the gamma-rays, and is dominated by
non-thermal radiation (synchrotron and inverse-Compton), likely emitted by
a relativistic jet pointed close to our line of sight. This so-called
"relativistic beaming" gives rise to many interesting effects, which
explain most blazar features \citep[see the Appendix of][]{up95}.  The
properties of misdirected blazars are consistent with those of radio
galaxies. Indeed, unified schemes \citep[e.g.,][]{up95} ascribe the
differences between "beamed"  (i.e., with their jets forming a small angle
w.r.t. the line of sight) objects and the so-called "parent population"
to orientation effects. Within the blazar class, which includes
flat-spectrum radio quasars (FSRQ) and BL~Lacertae objects, these are
thought to be the beamed counterparts of high- and low-luminosity radio
galaxies, respectively. The main difference between the two blazar classes
lies in their emission lines, which are strong and quasar-like for FSRQ and
weak or in some cases outright absent in BL~Lacs.

As a consequence of their peculiar orientation with respect to our line of
sight, blazars represent a rare class of sources, making up considerably
less than $5\%$ of all AGN \citep{pad97b}. Therefore, previously available
blazar samples suffered from small number statistics and relatively high
limiting fluxes (until recently $\sim $ 1 Jy and a few $\times 10^{-13}$
erg cm$^{-2}$ s$^{-1}$ in the radio and X-ray band respectively). The small
size of these samples ($\sim 30 - 50$ objects) implies also that the
derivation of beaming parameters based on luminosity function studies,
and therefore the viability of unified schemes \citep[e.g.,][]{pad90,ur91}
is considerably uncertain, especially at low powers. Moreover, as our
understanding of the blazar phenomenon is mostly based on relatively bright
and intrinsically luminous sources, we have only been sampling the tip of
the iceberg of the blazar population. For example, the best (and only!)
radio luminosity function (LF) of BL Lacs is still the 1 Jy one, which is
about 15 years old \citep{sti91}. The situation is only slightly better for
FSRQ. \cite{wal05} have recently studied the FSRQ LF and evolution using
the Parkes 0.25 Jy sample, while \cite{ri06} had to use the \cite{ku81} 1
Jy sample to study the epoch-dependency of the FSRQ LF. (The fact that
\cite{up95} used a 2 Jy sample for their review paper gives an idea of how
slow progress in this field is.)  This state of affairs, and the lack of a
sizeable sample which includes both FSRQ and BL Lacs, has so far also
prevented a specific test of the suggested possible evolutionary link
between FSRQ and BL Lacs \citep{cav02,bo02}.

The large majority of all known blazars have been discovered either in
radio or X-ray surveys \citep[but see][for a recent optically-selected BL
Lac sample identified from the Sloan Digital Sky Survey]{col05}. Previous
work has shown that X-ray and radio selection methods yield BL Lacs with
somewhat different properties, especially as regards the frequency at which
most of the synchrotron power is emitted, $\nu_{\rm peak}$
\citep{pad95b}. The spectral energy distributions (SEDs) of most radio
selected BL Lacs peak, in a $\nu - \nu f_{\nu}$ notation, in the IR/optical
bands, and these sources are now referred to as LBL \citep[low-energy
peaked BL Lacs:][]{pad95b}. By contrast, X-ray surveys select mostly HBL
(high-energy peaked BL Lacs), whose energy output peaks in the UV/X-ray
bands. The question of which of the two BL Lac subclasses is the most
numerous one has been a topic of debate in the past few years.  This is not
simply a demographical issue but touches upon the details of jet physics
and the so-called "blazar sequence" \citep[see below as well as][for a
review]{pad07}.  In this respect, we note that \cite{per98} and
\cite{pad02,pad03} have recently demonstrated that, contrary to previous
(lack of) evidence and the predictions of the blazar sequence, FSRQ with
SEDs similar to those of HBL, or HFSRQ, do indeed exist, although these
sources fail to reach $\nu_{\rm peak}$ values as extreme as those of
HBL. One way to answer this population question is through relatively deep
number counts.  However, BL Lac number counts are still based on samples
put together in the early nineties and, therefore, at relatively high
fluxes. Furthermore, while the strong evolution of FSRQ down to the (still
relatively high) fluxes sampled seems established, the issue of BL Lac
evolution is still quite open. X-ray selected samples (mostly HBL) have
generally shown indications of (small) negative evolution, i.e., with
sources being less luminous and/or less numerous in the past
\citep[see][and references therein; this result has been recently
challenged by \cite{cac02}]{rec00,bec03}. Radio-selected samples (mostly
LBL), on the other hand, have exhibited very weak, if any, evolution
\citep{sti91}. It is important to notice that no sample, so far, has
studied the evolutionary properties and LF of FSRQ, HBL, and LBL, within
the same survey.

Deeper, sizable blazar samples, have started to be assembled in the past
few years by us \citep[][hereafter Paper I and II respectively; see also
\cite{pad03}]{per98,lan01} and others \citep[RGB: \cite{lau99}; REX:
\cite{cac02}; HRX: \cite{bec03}; CLASS: \cite{cac04}; and Sedentary:][see
also \cite{pad02sax} for a review]{gio05}.  Some of these samples take
advantage of the fact that blazars are relatively strong radio and X-ray
sources and therefore use a double radio/X-ray selection method\footnote{We
put DXRBS and some of these surveys in perspective in \S~\ref{xrayradio}.}.

In Paper I and II we presented the methods, most of the identifications,
and some preliminary results of the Deep X-Ray Radio Blazar Survey (DXRBS),
a large-area ($\la 1,900$ deg$^2$ depending on X-ray flux) survey which
reaches relatively faint X-ray ($\sim 2 \times 10^{-14}$ erg cm$^{-2}$
s$^{-1}$) and radio ($\sim 50$ mJy, depending on declination) fluxes. In
this paper we present the radio number counts, evolutionary properties, and
luminosity functions of blazars in the DXRBS sample, which is at present
almost completely ($\ga 94\%$) identified.

There are various features that make DXRBS a unique sample with which
to address various issues:

\begin{enumerate}

\item DXRBS is currently the faintest and largest blazar sample with nearly
complete identifications; it therefore allows us to reach fainter and
intrinsically weak blazars, thereby providing further tests of unified
schemes;

\item DXRBS includes both FSRQ and BL Lacs {\it within the same sample}; it
is then possible to compare the properties of the two classes, including
number densities, luminosity functions, and evolution, independently of 
selection effects due to different sample criteria and definitions. 
This is vital also to test the idea of an evolutionary link
between the two classes;

\item DXRBS includes both LBL and HBL {\it within the same sample}; this
has never been achieved before and makes it possible to compare the
properties of the two classes independently of obvious effects
due to the different selection bands and methods, HBL being normally X-ray
selected and LBL being typically radio-selected;

\item DXRBS reaches radio (and X-ray) fluxes faint enough to provide
definite tests to the so-called "blazar sequence" \citep{fos98}, which
posits an inverse dependence of $\nu_{\rm peak}$ on intrinsic power. This
scenario not only envisions the non-existence of FSRQ with
SEDs similar to those of HBL, which are present
in DXRBS, but predicts also a dominance of HBL in X-ray and relatively
deep radio surveys, which, until now, could not be proven or disputed 
due to the lack of deep samples;

\end{enumerate}  

The DXRBS survey and its selection criteria, the sky coverage of the X-ray
catalogue we have used, WGACAT, and the sample selection and
identifications are described in \S~2. In \S~3 we analyze the DXRBS
completeness, while \S~4 describes the DXRBS number counts, and \S~5
discusses the sample evolutionary properties. We obtain the DXRBS
luminosity functions in \S~6,
%and discuss our results in \S~7. Finally, 
while \S~7 summarizes our conclusions.

Throughout this paper spectral indices are written $S_{\nu} \propto
\nu^{-\alpha}$ and the values $H_0 = 70$ km s$^{-1}$ Mpc$^{-1}$,
$\Omega_{\rm M} = 0.3$, and $\Omega_{\rm \Lambda} = 0.7$ have been
used \citep{spe03}. To compare some of our results with previous work
at times we have also adopted an $H_0 = 50$ km s$^{-1}$ Mpc$^{-1}$,
$\Omega_{\rm M} = 0$, and $\Omega_{\rm \Lambda} = 0$ (empty Universe)
cosmology.  Preliminary results on some of the topics addressed in
this paper were presented by \cite{pad01,pad02sax}.

\section{The DXRBS Sample}\label{survey}

The selection technique and identification procedures used for DXRBS have
been described in Paper I and II. We summarize here our final sample
selection, discuss the WGCAT sky coverage, and present our sample
definition and identifications.

\subsection{Candidate Selection}\label{selection}

DXRBS takes advantage of the fact that all blazars are relatively strong
X-ray and radio emitters. Selecting X-ray and radio sources with flat radio
spectra (one of the defining properties of the blazar class) is therefore a
very efficient way of finding these rare sources. By adopting a spectral
index cut $\alpha_{\rm r} \le 0.7$ DXRBS selects all FSRQ (defined by
$\alpha_{\rm r} \le 0.5$) and basically all BL Lacs, and excludes the large
majority of radio galaxies.

DXRBS initially was the result of a cross-correlation of all serendipitous
(i.e., excluding targets) X-ray sources in the publicly available {\it
ROSAT} database WGACAT95 \citep[first revision:][]{whi95}, having quality
flag $\ge 5$ (to avoid problematic detections), with a number of publicly
available radio catalogues (the chosen X-ray and radio catalogues were
selected on the basis of their large area and low flux limit). North of the
celestial equator, we used the 6 and 20 cm Green Bank survey catalogues GB6
and NORTH20CM \citep{gre96,whi92}, while south of the equator we used the 6
cm Parkes-MIT-NRAO catalogue PMN \citep{gri93}. For objects south of the
celestial equator, where a survey at a frequency different from the one of
the PMN (6 cm) was missing when we started this project (the NVSS
[\cite{con98}], now available, reaches in any case only $\delta =
-40^\circ$), we conducted a snapshot survey with the Australia Telescope
Compact Array (ATCA) at 3.6 and 6 cm. This not only gave us arcsecond radio
positions for our southern sources (we use the NVSS for the northern ones)
but also radio spectral indices unaffected by variability. Note that the
primary selection has been done at 6 cm, as the 20 cm catalogues are used
to derive spectral indices.

A second version of WGACAT was released in May 2000. In the course of the
work on this version, White, Giommi, and Angelini became aware of a problem
with the coordinates for 345 sequences in WGACAT95, caused by an error in
converting the header of the event files. This error caused an offset in
the source declination up to 1 arcmin but typically less. The fraction of
sources affected is very small, $\sim 1.4\%$ and $\sim 0.4\%$ for the PMN
and GB6 correlation respectively. For objects satisfying our completeness
criteria (see \S~\ref{definition}), no source was mistakenly included
because of this error, while only one had to be added. The $1\sigma$ 
WGACAT positional errors (see Paper I for more details) range from 13 
arcsec for the inner $10^{\prime}$ of the PSPC field to 53 arcsec for the 
$50 - 60^{\prime}$ ring.

The X-ray/radio matching was done as follows. WGACAT was correlated with
both GB6 and PMN catalogues with a radius of 1.5 arcmin, excluding however
sources for which the ratio between X-ray/radio offset and positional error
was larger than 2 for the inner region ($<30$ arcmin) of the {\it ROSAT}
Position Sensitive Proportional Counter (PSPC) (as for such relatively
large correlation radii one expects a non-negligible number of spurious
matches; see \S~\ref{cross}).  In the northern hemisphere the resulting
sample was then correlated with the NORTH20CM catalogue with a radius of 3
arcmin, as the positional uncertainties of the NORTH20CM catalogue are
considerably worse than those of the GB6 catalogue (160 arcsec at the 90\%
level compared to $10-15$ arcsec at the $1 \sigma$ level respectively), and
the $6-20$ cm spectral index calculated.  In the south we derived the
$6-20$ cm spectral index from PMN and NVSS data, summing up the flux from
all NVSS sources within 3 arcmin from the PMN position for $\delta >
-40^\circ$, while we used our own ATCA observation to derive the $3.6-6$ cm
spectral index for $\delta < -40^\circ$ (this slightly different criterion
translates in an estimated loss of only three blazar candidates: see
details in Paper II and also \S~\ref{definition}). Finally, we excluded
from our candidate list the following sources: clearly resolved out by our
ATCA observations, with $|b| \le 10^\circ$, within $5^\circ$ from the Large
and Small Magellanic Clouds (LMC, SMC) and M 31, and within $6^\circ$ from
the Orion Nebula.

\subsection{WGACAT Sky Coverage}\label{coverage}

The sensitivity of the {\it ROSAT} PSPC instrument, besides the obvious
dependence on exposure time and background intensity, is a strong function
of the position in the field of view. Consequently, the area of the sky
covered at any given flux (usually known as the sky coverage) is a complex
function of flux. Two basic factors are responsible for the off-axis radius
dependence: 1. the decrease of the instrument effective area at large
offset angles (vignetting effect); 2. the degradation of the Point Spread
Function (PSF) with distance from the center. Both factors can be described
analytically by the following relationships which give the minimum
detectable count rate $cr_{\rm min}$ in a PSPC image:

\begin{equation}\label{skycov1}
cr_{\rm min} = 900 ~ \sqrt{(b / t)} ~~~~~~~~~~~~~~~~~~~~~~~~~~~~~  r \leq 10'
\end{equation}
\begin{equation}\label{skycov2}
cr_{\rm min} = 900 ~ \sqrt{(b / t)} ~ \exp[(r-10) / 15] ~~~~~~~ r>10'
\end{equation}

\noindent
where $t$ is the exposure time in seconds, $r$ the off-axis radius
expressed in arc minutes, and $b$ the value of the local background
(counts/pixel/s). The dependence on $t$ and $b$, in background limited
exposures ($t \ga 2000 ~ s$), is given by the factor $\sqrt{(b / t)}$ while
the reduced sensitivity at large off-set angles due to vignetting and the
PSF degradation is described by the exponential term $\exp[(r-10) / 15]$
and is negligible at low ($r \leq 10^{\prime}$) off-axis angles.

The form of eqs. (\ref{skycov1}) and (\ref{skycov2}) and the value of its
parameters have been derived by studying the dependence of WGACAT source
count rates on $t$, $b$ and $r$. Eqs. (\ref{skycov1}) and (\ref{skycov2})
are somewhat conservative since some sources can still be detected just
below the threshold, but it ensures that the number of spurious sources and
source confusion are reduced, so that WGACAT can be used for statistical
studies.

The WGACAT sky coverage has then been computed simply inverting the
sensitivity laws (\ref{skycov1}) and (\ref{skycov2}) for each field. The
following areas of the PSPC field of view have been excluded from the
computation:
\begin{enumerate}
\item $r<1.8'$, to exclude the target of the PSPC observations;
\item $13'<r<24'$, to avoid the PSPC window structure which absorbs most of
the photons in this region due to the wobble motion applied to most
exposures;
\item $r>45'$, to take into account the strongly reduced PSPC sensitivity
at larger off-axis angles;
\item 10\% of the circular region between $r=24'$ and $r=45'$, to subtract
the area covered by the eight equally spaced ribs extending outward from
the inner ring.
\end{enumerate}

Finally, to take into account the DXRBS selection criteria (see
\S~\ref{cats}), the sky coverage has been calculated excluding the WGACAT
fields satisfying the following conditions: 
\begin{enumerate}
\item $\delta > 75^{\circ}$ 
\item $-10^{\circ} \le b \le 10^{\circ} $
\item circular regions around M 31, LMC, SMC ($R=5^{\circ}$) and the 
Orion Nebula ($R=6^{\circ}$)
\item (small) regions not covered by the PMN and the GB6 surveys.
\end{enumerate}

Count rates have been converted to $0.3-2$ keV fluxes assuming a power law
spectral model absorbed by an amount of neutral hydrogen equal to the
Galactic value as determined by the 21 cm measurements of \cite{dic90}.

%fig. 1
\begin{figure}%1
%\plotone{f1.eps}
\centerline{\includegraphics[width=9.5cm]{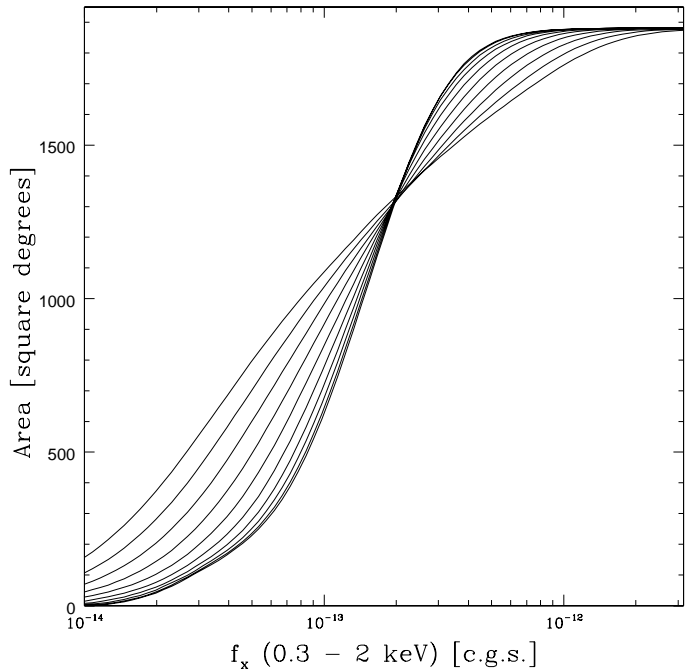}}
\caption{The DXRBS/WGACAT sky coverage for various
assumptions of the power law energy index, ranging from $\alpha_{\rm x}
=-0.2$ (bottom left curve) to $\alpha_{\rm x} =3.8$ (top left curve), with
a step equal to 0.4.\label{skycov}}
\end{figure}

Fig. \ref{skycov} shows the DXRBS/WGACAT sky coverage for various
assumptions of the power law energy index, ranging from $\alpha_{\rm x}
=-0.2$ to $\alpha_{\rm x} =3.8$, with a step equal to 0.4. Note how the
area covered at a given flux depends on the assumed energy slope. To
avoid the uncertainty introduced by assuming a single value, as often done,
this was estimated individually for every source from the hardness ratios,
as described in \cite{pad96}\footnote{Our results are basically unchanged
even assuming $\alpha_{\rm x} = 1$ or $\alpha_{\rm x} = 1.5$ for all our
sources.}.

\subsection{Sample definition and identifications} \label{definition}

We have defined a complete sample from the sources which meet
the following five criteria:

\begin{enumerate}
\item{$\alpha_{\rm 6-20} \leq 0.7$ for $\delta > -40^\circ$ and
$\alpha_{\rm 3.6-6} \leq 0.7$ for $\delta \le -40^\circ$;\footnote{This
slight difference in the wavelength range used to derive the radio spectral
index has very little effect on our sample. In Paper II, in fact, we showed
that the mean difference between $\alpha_{\rm 3.6-6}$ and $\alpha_{\rm
6-20}$ for core-dominated sources was 0.13. Expanding on the discussion
done there, we point out that there are no quasars with $\delta \le
-40^\circ$ and $0.5 < \alpha_{\rm 3.6-6} \leq 0.63$ satisfying our
completeness criteria, so basically no FSRQ is lost. Moreover, the fraction
of BL Lacs having $\alpha_{\rm r} > 0.7$ is very small ($\sim 5\%$ in the
\cite{pad97a} AGN catalogue).}}
\item{$|b| >10^\circ$, avoiding also the Large and Small Magellanic Clouds,
the Orion Nebula, and M 31;}
\item{$f_{20 {\rm cm}} \ge 150$ mJy for $0\le\delta\le75^\circ$;}
\item{$f_{6 {\rm cm}} \ge 51$ mJy for $-87.5^\circ \le \delta < -37^ \circ$
and $-29^\circ < \delta < 0^\circ$, $f_{6 {\rm cm}} \ge 72$ mJy for $-37^
\circ \le \delta \le -29^\circ$;}
\item{$1.8^\prime \le {\rm PSPC~offset} \le 13^\prime$ and $24^\prime \le
{\rm PSPC~offset} \le 45^\prime$.}
\end{enumerate}

Point 1 has been discussed in \S~\ref{selection}, point 2 excludes the
Galactic plane and the usual bright, extended sources, points 3 and 4
will be addressed in \S~\ref{complete}, while point 5 is related to the
derivation of the WGACAT sky coverage (see \S~\ref{coverage}). These
criteria were chosen in order to ensure that a well defined, flux-limited
sample could be obtained.

The above defining criteria were met by 219 blazar candidates, of which 77
were previously known sources. Details on the identification of the optical
counterpart and spectroscopic observations for most sources, including
objects not belonging to the complete sample, are given in Papers I and
II. Identifications for the remaining objects and the final list of DXRBS
sources will be presented in a future publication.  

We follow here a classification 
scheme slightly changed from that adopted in Paper II, using the work of
\cite{lan02,lan04}. The blazar class includes BL Lacertae objects, historically
characterized by an almost complete lack of emission lines, and the
flat-spectrum radio quasars (FSRQ), which by definition display broad,
strong emission lines. The separation of BL Lacs from radio galaxies, on
the one hand, and from radio quasars, on the other hand, is somewhat
controversial. Both BL Lacs and radio galaxies can have no or only narrow
emission lines and therefore, along the lines of \cite{ma96}, we have used
the value of the Ca break, a stellar absorption feature in the optical
spectrum defined by $C = (f_+ - f_-) / f_+$ (where $f_-$ and $f_+$ are the
fluxes in the rest frame wavelength regions $3750 - 3950$ \AA~ and $4050 -
4250$~\AA~ respectively), to differentiate between the two. We have adopted
a separation value of $C=0.4$, since \citet{lan02} showed that above this
value sources become increasingly lobe-dominated (see their Fig. 6).

We have adopted a dividing value of full-width half maximum $FWHM = 1000$
km/s between ``narrow'' and ``broad" emission lines. We also make the
commonly accepted distinction between steep spectrum radio quasars (SSRQ)
($\alpha_{\rm r} > 0.5$) and FSRQ ($\alpha_{\rm r} \le 0.5$). In order to
separate BL Lacs (with broad emission lines) from radio quasars we have
used the physical classification scheme of \citet{lan04} which separates
sources into weak-lined and strong-lined radio-loud AGN based on their
location in the rest frame equivalent width plane of the narrow emission
lines [OII] $\lambda 3727$ and [OIII] $\lambda 5007$ (see their Fig. 4). If
our spectrum did not cover the positions of both these emission lines we
have adopted for BL Lacs (i.e., weak-lined radio-loud AGN) a limit of 5
\AA~on the rest frame equivalent width of any detected emission line, which
is the same as that used for the 1 Jy sample \citep{sti91}.

Tab. \ref{tab1} gives the breakdown of the sample. 
One hundred and fifty-three sources turned out to be blazars, that is 129
FSRQ and 24 BL Lacs respectively. 

\begin{deluxetable}{lrrrr}
\tabletypesize{\scriptsize}
\tablecaption{Complete sample composition. \label{tab1}}
\tablewidth{0pt}
\tablehead{Class & Total & Newly & Previously & No redshift\\
                                 &  & identified & known &}
\startdata
FSRQ & 129 & 79 & 50  & 0\\
BL Lacs & 24 & 15 & 9 & 7\\
%BL Lacs, $\alpha_{\rm r} > 0.5$ & 7 & 6 & 1 & 0\\
SSRQ\tablenotemark{a} & 33 & 24 & 9 & 0\\
Radio Galaxies\tablenotemark{a}  & 17 & 9 & 8 & 0\\
Unidentified [$\alpha_{\rm r}  > 0.5$] & 16[6] & 16[6] & & 16[6]\\
%Unidentified,  $\alpha_{\rm r}  > 0.5$ & 6 & 6 & & 6\\
\enddata
\tablenotetext{a}{The SSRQ and radio galaxies samples are obviously
incomplete, as our radio spectral index cut excludes by definition the
majority of these sources.}
\end{deluxetable}

\section{Completeness of DXRBS}\label{complete}

The completeness of our sample is a complex function of many factors,
namely: 

\begin{enumerate}

\item the completeness of the catalogues which make up DXRBS; this is
clearly beyond our control but needs to be taken into account to the best
of our knowledge; 

\item our cross-correlation radii; one here has to find a
reasonable compromise between too large of a value, which produces a high
number of spurious sources, and too small of a value, which increases
incompleteness; 

\item our double X-ray/radio selection, which could result in some
incompleteness when comparing our results with purely radio-selected
samples;

\item our radio spectral index cut; 

\item the still missing identifications; 

\item a subtle second order effect, to do with non-serendipitous
sources being included, which could result in an excess of sources
at large radio fluxes.

%\item VARIABILITY ? 

\end{enumerate}

We discuss these points in turn below. 

\subsection{Input catalogues}\label{cats}

\subsubsection{GB6}

The GB6 catalogue \citep{gre96} covers the northern sky at $0^{\circ} <
\delta < +75^{\circ}$ with a flux limit at 6 cm which is declination
dependent and $\approx 25$ mJy. Since all our northern sources have 
$f_{\rm 20 cm} > 150$ mJy (see below), our spectral index cut of
$\alpha_{\rm r} \le 0.7$ implies $f_{\rm 6 cm} > 65$ mJy, well above the
GB6 limit. Indeed, all our northern sources have $f_{\rm 6 cm} > 87$ mJy.

\subsubsection{NORTH20}

The NORTH20 catalogue \citep{whi92} covers the sky at $-5^{\circ} < \delta
< +82^{\circ}$ with a formal flux limit at 20 cm of 100 mJy.  However, as
described in the original paper, the catalogue is known to be incomplete
below 150 mJy. We checked on this by using the NVSS. Out of the 32,530
sources in the NVSS with flux $\ge 100$ mJy, $-5^{\circ} < \delta <
+82^{\circ}$, and $|b| > 10^\circ$, only $\sim 72\%$ have a NORTH20
counterpart. For fluxes $\ge 150$ and 300 mJy this fraction goes up to
$\sim 90\%$ and $\sim 97\%$, respectively, which shows that there is still
some incompleteness above 150 mJy. To remedy this we cross-correlated the
WGACAT/GB6 sample with the NVSS using a radius of 1.5 arcmin. Spectral
indices were derived as described in Padovani et al. (2003), that is by
summing up the 1.4 GHz flux from all NVSS sources within a 3 arcmin radius
(corresponding roughly to the beam size of the GB6 survey). This resulted
in eleven sources with $f_{\rm 20 cm} > 150$ mJy, PSPC offset outside the
13 and 24 arcmin range (\S~\ref{definition}), and $\alpha_{\rm r} \le 0.7$
being added to our sample. We can then safely consider DXRBS complete for
$f_{\rm 20 cm} > 150$ mJy and $0^{\circ} < \delta < +75^{\circ}$ (the
region of the sky covered by both GB6 and NORTH20 catalogues).

\subsubsection{PMN}

The PMN catalogue \citep{gri93} comprises four different surveys in the
$-87.5^{\circ} < \delta < +10^{\circ}$ range, with flux limits at 6 cm
which are in most cases declination dependent and extend down to $\sim 20$
mJy. As the GB6 catalogue reaches $\delta = 0^{\circ}$, we did not include
any PMN ``northern'' source. Also, for simplicity we included in our
cross-correlation only sources with $f_{\rm 6 cm} > 51$ mJy in the area
covered by the Southern, Tropical, and Equatorial surveys ($-87.5^{\circ} <
\delta < -37^{\circ}$ and $-29^{\circ} < \delta < +10^{\circ}$), which is
the {\it maximum} declination-dependent flux limit for these surveys, and
sources in the Zenith survey ($-37^{\circ} < \delta < -29^{\circ}$) with
$f_{\rm 6 cm} > 72$ mJy, which is its limit. Apart from the declinations
covered by the Zenith survey, we are then well above the formal
completeness limits of the PMN survey.

\subsubsection{WGACAT}

The first version of WGACAT \citep{whi95} covers $\sim 10\%$ of
the sky to varying degrees of sensitivity, with exposures typically a
factor of 100 longer than those achieved during the six month {\it ROSAT}
All Sky Survey (RASS). Since we cross-correlate WGACAT with radio 
catalogues, the chance that one of our sources is a spurious X-ray detection 
is vanishingly small. The DXRBS X-ray flux limits depend on the exposure time and the
distance from the center of the PSPC (see \S~\ref{coverage}) and vary
between $\sim 10^{-14}$ and $\sim 10^{-11}$ erg cm$^{-2}$ s$^{-1}$.

\subsection{Cross-correlations}\label{cross}

A cross-correlation between two catalogues will produce a number of
spurious associations $N_{\rm sp}$ which depends on the number of sources
in the first (X-ray) catalogue $N_1$, the surface density in the second
(radio) catalogue $S$, and the correlation radius $r$ as follows

\begin{equation}\label{sp}
N_{\rm sp} = 8.73 \times 10^{-4} (r/arcmin)^2~(S/[\#/deg^2])~N_1.
\end{equation}

The number of spurious matches in a given thin shell of radius $\Delta r$
will be

\begin{equation}\label{spdelta}
N_{\rm sp}/\Delta r = 1.75 \times 10^{-3} (r/arcmin)~(S/[\#/deg^2])~N_1.
\end{equation}

\begin{figure}%2
%\plotone{f2.eps}
\centerline{\includegraphics[width=9.5cm]{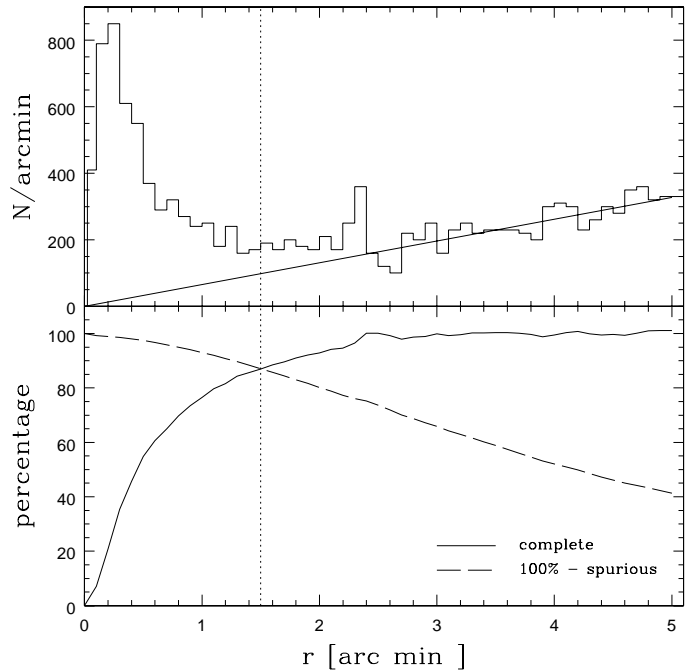}}
\caption{{\it Top}: the number of matches per arcmin shell as a function of
positional offset for the WGACAT -- GB6 correlation. The strong peak at
small radii is due to real matches, while the linear increase at large
radii is due to spurious sources. The solid line represents the number of
spurious sources as predicted by eq. (\ref{spdelta}), fitted by assuming
that at large radii all sources are spurious. {\it Bottom}: the dependence of
completeness (solid line) and real matches ($1~-$ spurious) fraction
(dashed line) on offset (see text for details). The dotted line at 1.5
arcmin defines our adopted cross-correlation radius. \label{completeness}}
\end{figure}

The completeness of our cross-correlation was computed in a way which is as
independent as possible of ``a priori'' assumptions, including the
positional errors, but relies only on the previous equations. Namely, the
cross-correlations were done with a radius larger than the one we finally
adopted (typically 5 arcmin) and the matches were binned in radial shells
(typically 6 arcsec) as $N/\Delta r$. The results were then plotted as a
function of the positional offset between the matches. This is shown for
the WGACAT -- GB6 correlation in Fig. \ref{completeness} (top), which
displays the strong peak at small radii due to real matches, followed by a
decline, and then by the rise $\propto r$ due to the predominance of
spurious sources at large radii.

The total number of spurious sources for a given correlation radius could
be estimated using eq. (\ref{sp}). However, this requires a reliable
estimate of $S$, the surface density of sources in the radio catalogue,
which might not be straightforward to determine when the flux limit depends
on position, as is the case for both GB6 and PMN catalogues. Most
importantly, the coefficients in eq. (\ref{sp}) and (\ref{spdelta}) neglect
clustering, which would slightly increase $N_{\rm sp}$. Therefore, we left
the numerical coefficient free and found the best-fit value by imposing
that above a large enough radius (determined interactively) all sources are
spurious, that is $N_{\rm sp}/\Delta r = N/\Delta r$ (solid line in
Fig. \ref{completeness} [top]). This gives us also an estimate of the
completeness level at every correlation radius, which is defined as $N_{\rm
real}(r)/N_{\rm real}(r \rightarrow \infty)$, where $N_{\rm real}(r) = N(r)
- N_{\rm sp}(r)$ (in practice $N_{\rm real}(r)$ converges at $r \ga$ 3
arcmin).

%fig. 2

Fig. \ref{completeness} (bottom) shows the dependence of completeness and
real matches ($1~-$ spurious) fraction as a function of positional offset
for the WGACAT -- GB6 correlation. We adopted as our correlation radius the
value at which the two curves cross, that is 1.5 arcmin (the same result
applies to the WGACAT -- PMN correlation). That is, we are maximizing our
completeness while at the same time minimizing the fraction of spurious
sources. Our results are as follows: the completeness level is $\sim 87\%$
for the WGACAT -- GB6 correlation and $\sim 85\%$ for the WGACAT -- PMN
correlation. The percentages of spurious matches are $\sim 13\%$ and $\sim
17\%$ for the GB6 and PMN correlations respectively, very close to the
fraction of sources with ratio between X-ray/radio offset and positional
error larger than 2 ($\sim 13\%$ and $\sim 21\%$), which were already
excluded from the sample (\S~\ref{selection}). For our adopted
cross-correlation radius of 3 arcmin, the completeness level of the
WGACAT/GB6 -- NORTH20CM correlation is $\sim 99\%$ (this very high value is
explained by the fact that we are correlating two radio catalogues at
similar frequencies).  As discussed above, the positional uncertainties of
WGACAT sources depend on their PSPC offset, so one might worry that in the
outer parts we could be more incomplete (see a preliminary discussion in
Paper I). However, the completeness level for the WGACAT -- PMN correlation
for PSPC offsets between 30 and 45 arcmin is $\sim 83\%$, not significantly
different from that of the full WGACAT ($\sim 85\%$).

To take the incompleteness related to the adopted cross-correlation radius
into account all our number counts and luminosity functions have been
multiplied by the factor $1/0.85 \sim 1.18$.

\subsection{The double X-ray/radio selection}\label{xrayradio}

%fig. 3
\begin{figure}%3
%\plotone{f3.eps}
\centerline{\includegraphics[width=9.5cm]{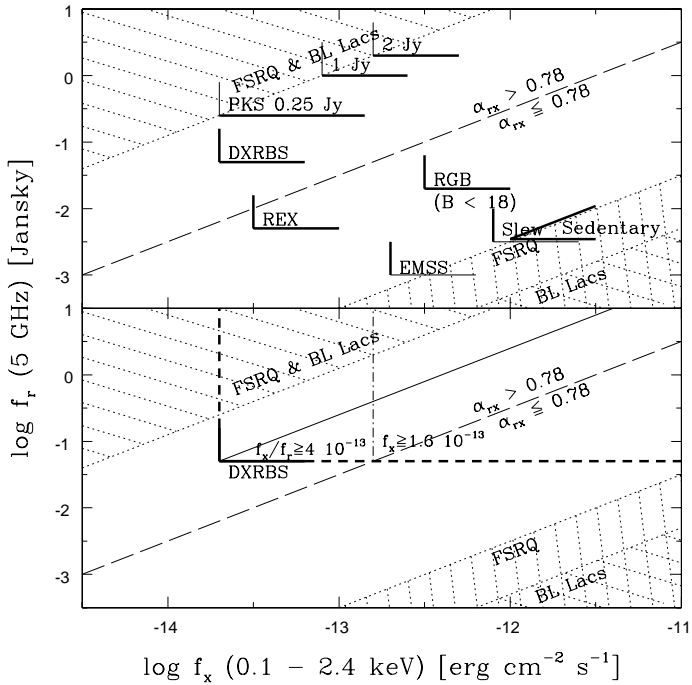}}
\caption{{\it Top}: The sampling of the radio flux -- X-ray flux plane by
different blazar surveys. Thick lines represent "hard'' survey limits,
while thin lines are the fluxes reached in a band other than the one of
selection. Sources belonging to a given survey occupy a region of the plane
whose bottom-left corner is indicated by the thick/thin lines. The
long-dashed line at $f_{\rm x}/f_{\rm r} = 3.2 \times 10^{-12}$ erg
cm$^{-2}$ s$^{-1}$ Jy$^{-1}$ separates sources with $\alpha_{\rm rx} \le
0.78$ (high-energy peaked blazars) from sources with $\alpha_{\rm rx} >
0.78$ (low-energy peaked blazars), while the hatched regions represent the
"forbidden'' zones, different for BL Lacs and FSRQ at the high $f_{\rm
x}/f_{\rm r}$ end, where no blazars have been found so far. {\it Bottom}: The
short-dashed lines show the parameter space sampled by DXRBS. The diagonal
solid line defines a DXRBS subsample with complete coverage of the plane
for $f_{\rm x}/f_{\rm r} \ge 4 \times 10^{-13}$ erg cm$^{-2}$ s$^{-1}$
Jy$^{-1}$, while the vertical dot-dashed line defines a complete X-ray flux
limited DXRBS LBL sample with $f_{\rm x} \ge 1.6 \times 10^{-13}$ erg
cm$^{-2}$ s$^{-1}$. See text for more details.\label{fxfr}}
\end{figure}

As many recent blazar surveys, DXRBS is not simply radio or X-ray
flux-limited, like the previous ``classical'' blazar samples, but adopts a
double radio/X-ray selection. As discussed by \cite{pad01,pad02sax}, it is
important to assess what region of parameter space such a survey is
sensitive to, in order to understand what constraints it can or cannot put
on blazar demographics. In particular, one should ask if such a sample can
provide meaningful radio or X-ray constraints (number counts, luminosity
functions, etc.), i.e., if it can provide a representative blazar sample
which can be considered flux-limited in one band. The answer to this lies
in how a survey covers the blazar radio flux -- X-ray flux plane and how 
this compares to the region of this plane which can be occupied by blazars.

To this aim, we have estimated the range of $f_{\rm x}/f_{\rm r}$ covered
by blazars using the multiwavelength AGN catalogue put together by
\cite{pad97a}, adding additional X-ray information from \cite{sie98}.  We
get a value for the smallest X-ray-to-radio flux ratio $\sim 8 \times
10^{-14}$ erg cm$^{-2}$ s$^{-1}$ Jy$^{-1}$, for both FSRQ and BL Lacs
(upper diagonal dotted line in Fig. \ref{fxfr}). As regards the largest
value, this differs by about one order of magnitude for the two classes, as
expected based on the results of \cite{pad03}, being $f_{\rm x}/f_{\rm r}
\sim 2 \times 10^{-9}$ erg cm$^{-2}$ s$^{-1}$ Jy$^{-1}$ for BL Lacs and
$f_{\rm x}/f_{\rm r} \sim 3 \times 10^{-10}$ erg cm$^{-2}$ s$^{-1}$
Jy$^{-1}$ for FSRQ (lower diagonal dotted lines in Fig. \ref{fxfr}). This
agrees with the values found by \cite{gio06} by correlating the NVSS
sources with $f_{\rm r} > 1$ Jy with the RASS.

These {\it ROSAT} based results hold even at fainter X-ray fluxes.
\cite{bas04} present X-ray data for 15 high-redshift radio-loud quasars, 12
of which have radio spectral index information and classify as FSRQ
(this is by no means a complete sample, although
it represents $\sim 30\%$ of all currently known $z > 4$ radio-loud
quasars). The minimum $f_{\rm x}/f_{\rm r}$ value for this sample is $\sim
10^{-13}$ erg cm$^{-2}$ s$^{-1}$ Jy$^{-1}$, i.e., basically the same as the
one obtained based on {\it ROSAT} data, despite the fact that the {\it
typical} X-ray flux for this sample is $\sim 10^{-13}$ erg cm$^{-2}$
s$^{-1}$, i.e., about one order of magnitude smaller than for DXRBS.

Fig. \ref{fxfr}\footnote{This is an updated and revised version of a figure
originally shown in \cite{pad01,pad02sax}.} (top) shows the sampling of the
radio flux -- X-ray flux plane by DXRBS compared to that by the
``classical'' blazar surveys (1 Jy \citep{sti91}, {\it Einstein} Medium
Sensitivity Survey (EMSS) \citep{sto91,rec00}, and {\it Einstein} Imaging
Proportional Counter (IPC) Slew \citep{per96} BL Lac samples and the 2 Jy
FSRQ sample \citep{wal85,ser94}) and a number of recent surveys with double
radio/X-ray selection (RASS - Green Bank (RGB) survey \citep{lau99}, the
Radio-Emitting X-ray (REX) survey \citep{cac99,cac02}, and the Sedentary
survey \citep{gio99,gio05}). Note that all of these recent surveys are
primarily devoted to BL Lacs, although REX includes also emission line
sources and \cite{pad03} have cross-correlated the RGB sample with the NVSS
to obtain radio spectral indices and extract the FSRQ. We also add a very
recent FSRQ survey which, although not as deep as DXRBS and purely
radio-selected, goes deeper than the 2 Jy sample, namely the Parkes 0.25 Jy
flat-spectrum sample \citep{wal05}.  To the best of our knowledge, all
other recent surveys which reach fainter radio fluxes than DXRBS are either
not completely identified and/or have a relatively bright optical magnitude
cut which translates into severe constraints on the number and type of
sources included.

In Fig. \ref{fxfr} (top) every survey is characterized by either one or two
flux limits (thick lines), while the smallest flux reached by a sample in a
band other than the one of selection is given by a thin line. For example,
the EMSS reaches $f_{\rm x} \sim 2 \times 10^{-13}$ erg cm$^{-2}$ s$^{-1}$
(thick line), by default the X-ray selection limit (actually, the faintest
of various X-ray limits, due to the serendipitous nature of the survey). A
limit in one band translates into a limit in the other one and in this case
the radio faintest EMSS BL~Lac has a flux $f_{\rm r} \sim 1$~mJy (thin
line). The sources of a given survey occupy a region of the flux-flux plane
whose bottom-left corner is shown in the figure. The long-dashed line in
the figure (X-ray-to-radio flux ratio $f_{\rm x}/f_{\rm r} = 10^{-11.5}$
erg cm$^{-2}$ s$^{-1}$ Jy$^{-1}$ or $\alpha_{\rm rx} \sim 0.78$) divides
high-energy peaked from low-energy peaked blazars \citep[both BL Lacs and
FSRQ; see][]{pad02,pad03}

Fig. \ref{fxfr} clearly illustrates the limitations of surveys with double
radio/X-ray flux limits. A survey whose limits fall quite far from both
dotted lines will not provide a complete picture of the blazar
population. For example, the REX survey cannot provide BL~Lac radio number
counts to be compared with the predictions of a beaming model based on the
1~Jy sample, simply because it does not include all the BL~Lacs above its
radio limit (as it misses all those above the radio limit but below the
X-ray limit). For the complementary reason, neither can REX provide X-ray
number counts to be compared with the predictions from a beaming model
based on the EMSS sample. REX will provide radio number counts for HBL,
given its proximity to the HBL/LBL dividing line, and X-ray number counts
for LBL (as it detects all LBL above its X-ray flux limit). The same
arguments apply to the RGB survey, which has the further problem of an
optical limit ($B < 18$). In short, {\it surveys with two flux limits can
provide a complete picture of the blazar population only if one of the
limits is relatively close to the edge of the region of parameter space
occupied by blazars\/}.

DXRBS misses sources only in a small corner of the radio flux -- X-ray flux
plane between its X-ray flux limit and the lower limiting X-ray-to-radio
flux ratio for blazars (Fig. \ref{fxfr}, bottom). For a value of this
parameter $\sim 8 \times 10^{-14}$ erg cm$^{-2}$ s$^{-1}$ Jy$^{-1}$ and a
radio limit of 51 mJy, in fact, DXRBS should have been sensitive to X-ray
fluxes down to $\sim 4 \times 10^{-15}$ erg cm$^{-2}$ s$^{-1}$ to be
considered equivalent to a purely radio flux-limited sample. It then
follows that no source with $51 \le f_{\rm 6 cm} < 250$ mJy and $f_{\rm x}
< 2 \times 10^{-14}$ erg cm$^{-2}$ s$^{-1}$ (see Fig. \ref{fxfr}) could
have been included in the survey. In fact, DXRBS covers fully the blazar
region of the radio flux -- X-ray flux plane at radio fluxes $\ge 250$ mJy,
as for this radio limit and given DXRBS' X-ray limit, our survey touches
the "forbidden" zone in Fig. \ref{fxfr} (and happens to overlap with the
PKS 0.25 Jy). Furthermore, DXRBS includes {\it all} blazars with $f_{\rm r}
\ge 51$ mJy and $f_{\rm x}/f_{\rm r} \ge 4 \times 10^{-13}$ erg cm$^{-2}$
s$^{-1}$ Jy$^{-1}$, as shown by the solid line in Fig. \ref{fxfr}, bottom. 
We will address this point again in the following sections, but we
stress here that the missed small corner of the radio flux -- X-ray flux
plane has no influence on the number counts, evolution, and luminosity
function of DXRBS, since to evaluate these we properly take into account
our flux limits. However, a correction is required when comparing these
properties to those of purely radio-selected samples (see \S~\ref{counts}). 
Finally, an X-ray
flux limited LBL sample with $f_{\rm x} \ga 1.6 \times 10^{-13}$ erg
cm$^{-2}$ s$^{-1}$ (dot-dashed line in Fig. \ref{fxfr}, bottom) can be
defined within DXRBS (see Section \ref{relative}).

\subsection{Radio spectral index cut}\label{radiocut}

Our sources are selected to have $\alpha_{\rm r} \le 0.7$
(\S~\ref{definition}).  However, the radio data used to derive the radio
spectral index are simultaneous only for $\delta \le
-40^{\circ}$. \cite{wal05} have recently addressed this issue and pointed out that
any flux-limited survey will preferentially select sources in an up-state,
whereas flux density measurements taken at a different time reflect sources
in a mean state. They also point out that it is the variations at
frequencies {\it above} the survey frequency which matter. Since our radio
spectral indices have been derived using frequencies {\it below} the survey
frequency, this would result in an artificial flattening of the spectra,
with the inclusion of some extra sources in our sample. However, the fact
that all our data come from flux-limited surveys is bound to mitigate this
effect.

\subsection{Missing identifications}\label{unident}

Our sample is not yet fully identified. At the time of writing (January 2007),
there are still sixteen sources in the complete sample for which we do not
have a classification (seven in the southern and nine in the northern
hemisphere). Out of the 16 still unidentified objects, we note
that six have $\alpha_{\rm r} > 0.5$ and therefore cannot be FSRQ. Based on
the relative fraction of the different DXRBS classes, we estimate that we
are missing $\approx 9$ FSRQ and $\approx 2$ BL Lacs, with the remaining
five sources being SSRQ (4) and radio-galaxies (1). Therefore, we are
likely to still be missing $\approx 7\%$ and $\approx 8\%$ of DXRBS FSRQ
and BL Lacs respectively. We address this incompleteness in
\S~\ref{evolution}.

\subsection{Non-serendipitous sources}\label{serend}

Although we have obviously excluded all {\it ROSAT} targets from our
samples, a subtle, second-order effect is present which results in the
inclusion of additional non-serendipitous sources.  Consider the case of an
X-ray target which was discovered because in the field of view of a
well-known and, therefore, likely radio-bright blazar. When this source is
observed as a {\it ROSAT} target, the quasar will appear to be
serendipitously in its field of view, while in fact it is not there by
chance and needs to be excluded. We have then done so for the easiest cases
(e.g., radio-loud objects in fields of {\it Einstein} Medium Sensitivity
Survey sources), also by going back to the original {\it ROSAT} proposal,
but in some other cases it is very difficult to ascertain the relation
between the target and a radio source in the field. This might translate
into a residual excess at high radio fluxes (see \S~\ref{counts}).

%\subsection{Variability}\label{variab}

\section{Number Counts}\label{counts}

The area of the sky over which every source could be found, used for the number
counts, volume calculations, and LFs, is determined by its X-ray and radio
fluxes as follows:

\begin{enumerate}

\item if $f_{6 {\rm cm}} \ge 72$ mJy and $f_{20 {\rm cm}} \ge 150$ mJy, the
radio source is detectable over the whole sky; we then use the full WGACAT
sky coverage and determine the appropriate area based on the values of the
X-ray flux and X-ray spectral index;

\item if $51 \le f_{6 {\rm cm}} < 72$ mJy and $f_{20 {\rm cm}} \ge 150$ mJy,
the radio source is detectable over the whole sky excluding the PMN
Zenith area ($-37^{\circ} < \delta < -29^{\circ}$); we then use the WGACAT
sky coverage for the whole sky excluding that region;
 
\item if $f_{6 {\rm cm}} > 72$ mJy and $f_{20 {\rm cm}} < 150$ mJy, the
radio source is not detectable in the northern hemisphere; we then
use the WGACAT sky coverage for the southern sky;

\item if $51 \le f_{6 {\rm cm}} < 72$ mJy and $f_{20 {\rm cm}} < 150$ mJy,
the radio source is detectable over the whole southern sky excluding
the PMN Zenith area ($-37^{\circ} < \delta < -29^{\circ}$); we then use the
WGACAT sky coverage for the southern sky excluding that region.

\end{enumerate}

As discussed in \S~\ref{xrayradio}, DXRBS is "almost" equivalent to a radio
flux-limited sample. In fact, it misses some (but not all) sources with $51
\le f_{\rm 6 cm} < 250$ mJy and $f_{\rm x}/f_{\rm r} < 4 \times 10^{-13}$
erg cm$^{-2}$ s$^{-1}$ Jy$^{-1}$ (see Fig. \ref{fxfr}). We corrected the
number counts for this effect by first deriving the $f_{\rm x}/f_{\rm r}$
distribution for the sub-sample with $f_{\rm 6 cm} > 250$ mJy
\citep[weighted for the effect of the sky coverage; see][]{pad03}. This
sub-sample covers the full range of $f_{\rm x}/f_{\rm r}$ for blazars and
therefore its $f_{\rm x}/f_{\rm r}$ distribution should be unbiased. We
then compared that to the $f_{\rm x}/f_{\rm r}$ distribution for the whole
sample and estimated the fraction of missed sources with $f_{\rm x}/f_{\rm
r} < 4 \times 10^{-13}$ erg cm$^{-2}$ s$^{-1}$ Jy$^{-1}$ as a function of
$f_{\rm x}/f_{\rm r}$. The $f_{\rm x}/f_{\rm r}$ values were then converted
to radio fluxes by using our X-ray limit and the number counts were then
corrected. (For example, the correction for $f_{\rm 6 cm} > 100$ mJy was
derived by evaluating the fraction of additional sources with $f_{\rm
x}/f_{\rm r} < 2 \times 10^{-13}$ erg cm$^{-2}$ s$^{-1}$ Jy$^{-1}$, this being
equal to our X-ray flux limit of $2 \times 10^{-14}$ erg cm$^{-2}$
s$^{-1}$ divided by this radio flux.) The correction is not large, being
$\sim 13\%$ for $100 < f_{\rm 6 cm} < 250$ mJy and $\sim 70\%$ only for
$f_{\rm 6 cm} < 100$ mJy, and within $\approx 1\sigma$ from the uncorrected
values (compare the dotted-dashed and solid lines in Fig. \ref{blcounts}
and \ref{fscounts}). Given the small number statistics for the BL Lac
sample (only 8 BL Lacs have $f_{\rm 6 cm} > 250$ mJy), we evaluated this
for FSRQ and then applied the same correction to the BL Lac sample.

\subsection{BL Lacs} 

%fig. 4
\begin{figure}%4
%\plotone{f4.eps}
\centerline{\includegraphics[width=9.5cm]{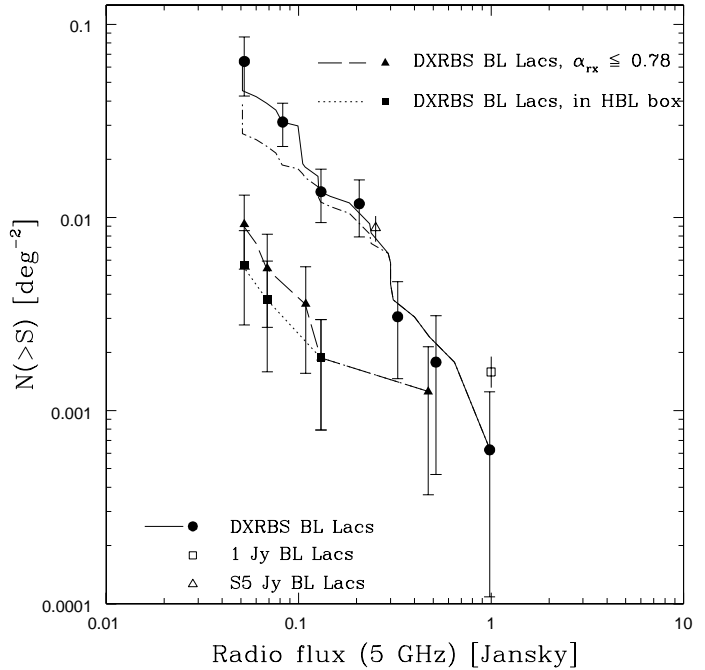}}
\caption{The integral number counts for DXRBS BL Lacs at 5 GHz (6 cm). The
solid line represents the total number counts, while filled circles show
the values at selected fluxes to show the errors involved. The
dotted-dashed line shows the counts without the correction which we applied
to make DXRBS equivalent to a radio flux-limited sample
(\S~\ref{xrayradio}). The open square represents the surface density for
the 1 Jy sample, while the open triangle represents the surface density for
the S5 sample. Both of these points have been corrected to take into
account our somewhat different definition of a BL Lac. The dashed line and
filled triangles are the number counts for the BL Lacs with $\alpha_{\rm
rx} \le 0.78$, while the dotted line and filled squares are the number
counts for the BL Lacs in the HBL box. See text for
details.\label{blcounts}}
\end{figure}

Fig. \ref{blcounts} presents the integral number counts at 5 GHz for the
DXRBS BL Lacs (solid line), compared to the values derived from the 1 Jy
\citep[][open square]{sti91} and S5 \citep[][as updated by \cite{sti93};
open triangle]{ku87} samples. The filled circles show the values at
selected bins to show the errors involved without crowding the plot.  In
making the comparison with previous BL Lac surveys we note that their
definition of BL Lac was somewhat different from ours, which is less
restrictive and reaches $\alpha_{\rm r} = 0.7$ (while previous radio
samples were defined by $\alpha_{\rm r} \le 0.5$). Since $2/3$ (16/24) of
the DXRBS BL Lacs also fulfill the 1 Jy definition, the 1 Jy and S5 points
were multiplied by the factor $3/2$ to take this difference into account.

A couple of points need to be considered before discussing our results:
1. we do not characterize the uncertainties by simply considering the total
number of sources, as the error bars would become progressively smaller,
which is misleading. At relatively faint X-ray (and, on average, radio)
fluxes, in fact, the surveyed area decreases and the number of sources gets
smaller. We take that into account by summing in quadrature the errors for
the individual sources. As a result, the error bars are larger at high and
low fluxes, where the number of sources is small, and smaller at
intermediate fluxes, where the statistics is better; 2. the radio flux
limit is higher at 20 cm than at 6 cm.  This has an impact on the northern
counts only below 150 mJy, our adopted completeness limit at 20 cm, as for
$\alpha_{\rm r} \sim 0$ (the median for our sample), $f_{\rm 6 cm} \sim
f_{\rm 20 cm}$. Given the lower flux limit of the 6 cm (PMN) survey, the
southern counts go deeper than the northern ones.

Taking all this into account, we can say that our counts agree with
previous estimates at relatively high fluxes (250 mJy). The BL Lac with the
largest radio flux in our sample has $f_{\rm 6 cm} \sim 0.99$ Jy so we
cannot really compare our results with the 1 Jy sample. Given the fact
that for the northern and southern hemisphere we have used different radio
catalogues and a somewhat different selection, we have also checked that
the southern and northern counts agree within the errors down to $\sim 150$
mJy; below this value only southern sources contribute.

\subsection{The relative numbers of HBL and LBL}\label{relative}

Until about ten years ago HBL (then called XBL for X-ray selected BL Lacs)
were regarded to have their jets seen at larger angles w.r.t. to the line
of sight as compared to LBL \citep[then called RBL for radio-selected BL
Lacs; see, e.g.,][and references therein]{up95}. This meant also that
XBL/HBL were thought to be the most numerous subclass. In the mid 1990s
\cite{gio94} and \cite{pad95a} proposed a new interpretation for the
existence of HBL/XBL. Against the prevailing view at the time, HBL/XBL were
instead suggested to be intrinsically rare and to represent the $\approx
10\%$ of BL Lacs with relatively high $\nu_{\rm peak}$. The fact that they
were the dominant class in the X-ray band was thought to be a simple
selection effect related to their SED. Namely, X-ray surveys were sampling
the BL Lac radio population at relatively low radio fluxes and mostly
detected the small fraction of objects with high $f_{\rm x}/f_{\rm r}$
ratios. This scenario is now also supported by the results of \cite{lan02},
which show no difference in jet orientation between the two BL Lac
subclasses \citep[see also][]{rec00}.

By the end of the 1990s the ``different orientation'' paradigm was replaced
by a new scenario. The so-called ``blazar sequence''
proposed that LBL and HBL sampled the higher and lower part of the jet
bolometric luminosity function, respectively, instead of different parts of
the radio luminosity function \citep{fos97}. The ``blazar sequence'', which
was later expanded to incorporate also radio quasars \citep{fos98},
advocates once more that HBL are more numerous than LBL. Furthermore, since
it is based on the assumption that an inverse dependence exists between
$\nu_{\rm peak}$ and intrinsic power due to the effects of the more severe
electron cooling in more powerful sources, it excludes the existence of
blazars with high radio powers and high $\nu_{\rm peak}$ (i.e., HFSRQ).

The predictions for the relative numbers of HBL and LBL in radio and X-ray
surveys are dramatically different in the two scenarios and can be put to
test with deep blazar surveys. In the first case \citep{pad95a}, HBL
represent a minority, and their fraction in radio surveys should be
constant and $\approx 10\%$. Moreover, X-ray surveys should detect HBL in
large numbers at high fluxes, due to their UV/X-ray peaked SEDs, but deeper
X-ray samples should reveal an increasingly large fraction of LBL
\citep{gio94,pad95a}. In the second case the situation is reversed, with
the fraction of HBL expected to increase at lower fluxes in the radio band
and be basically constant in the X-rays \citep{fos97}.

Note that the question of which BL Lac class (HBL or LBL) is most numerous
is not simply a demographical question but has also strong implications on
the physics of relativistic jets. The frequency at which most of the
synchrotron power is emitted, $\nu_{\rm peak}$, in fact, shows a very large
range in BL Lacs. Furthermore, $\nu_{\rm peak} \propto \gamma^2_{\rm peak}
\delta B$, where $\gamma_{\rm peak}$ is the Lorentz factor of the electrons
emitting most of the radiation, $\delta$ is the Doppler factor, and $B$ is
the magnetic field. Evidence towards a predominance of LBL or HBL would
then point towards Nature preferentially making jets which peak at
IR/optical or UV/X-ray energies, thereby constraining also the physical
parameters of the jets.

In order to test the two competing scenarios we have also derived the
number counts for HBL. HBL should be defined in terms of the position of
the frequency at which most of the synchrotron power is emitted, $\nu_{\rm
peak}$. This requires multifrequency data and fitting their SEDs. One
alternative is to use the X-ray-to-radio flux ratio, or the effective
radio-X-ray spectral index $\alpha_{\rm rx}$. Bona fide HBL, in fact, were
shown to have $\alpha_{\rm rx} \le 0.78$ \citep{pad96}. \cite{pad03},
however, have found that a definition based solely on $\alpha_{\rm rx}$ was
not optimal, while the position of the sources on the $\alpha_{\rm ox}$,
$\alpha_{\rm ro}$ plane, which means using two (instead of one) effective
spectral indices, appeared to be more sensitive to the synchrotron peak
frequency, especially for FSRQ. They then defined a so-called ``HBL box'',
derived by using all HBL in the multi-frequency AGN catalog of
\citet{pad97a}, i.e., a region of this plane within $2\sigma$ from the mean
$\alpha_{\rm ro}$, $\alpha_{\rm ox}$, and $\alpha_{\rm rx}$ values of
HBL. For comparison with previous results we adopt in this paper both
definitions.

The situation in the radio band is illustrated in Fig. \ref{blcounts}. The
fraction of BL Lacs with $\alpha_{\rm rx} \le 0.78$ is $14^{+9}_{-6}\%$,
while that in the HBL box is $9^{+6}_{-3}\%$. These values are slightly
smaller than those given in \cite{pad03} due to our somewhat different BL
Lac definition and also because in that paper we did not correct for the
missing corner in the radio flux -- X-ray flux plane discussed above and in
\S~\ref{xrayradio} (Note that no correction is required for HBL sources, as
their X-ray fluxes are such that they are all well away from that corner:
see Fig. \ref{fxfr}). Interestingly, the fraction of HBL, according to both
definitions, appears to be constant with radio flux, within the errors, at
least where we have enough statistics ($f_{\rm r} \la 0.1$ Jy). Said
differently, the ratio LBL/HBL is constant and $\sim 6$ (using the
$\alpha_{\rm rx}$ definition, for consistency with previous work). At the
radio fluxes reached by DXRBS, \cite{fos97} predicted a value $\sim 2.1$
(extrapolating from their Tab. 3), that is a factor of 3 smaller.
Fig. \ref{fscounts} (see \S~\ref{fscounts_sect}) paints a strikingly
similar picture for FSRQ.

%fig. 5
\begin{figure}%5
%\plotone{f5.eps}
\centerline{\includegraphics[width=9.5cm]{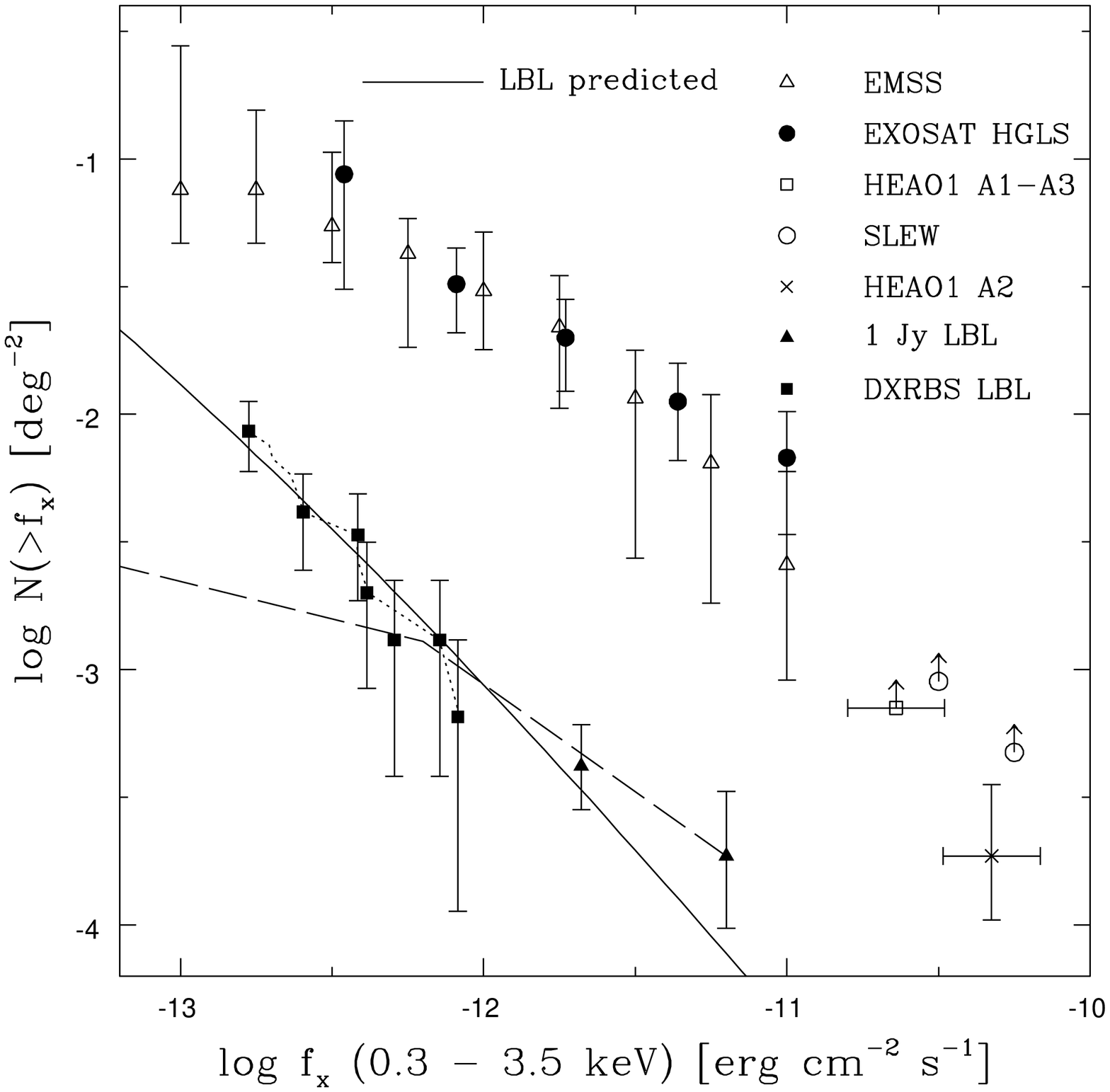}}
\caption{The integral X-ray number counts for BL Lacs, adapted from
\cite{pad95a}. Data for five X-ray selected samples are shown
\citep[see][for details]{pad95a}. Filled triangles represent the bivariate
X-ray counts for the 1 Jy LBL with $f_{\rm x} \ga 3 \times 10^{-12}$ erg
cm$^{-2}$ s$^{-1}$. The dotted line represents the DXRBS LBL with $f_{\rm
x} \ge 1.6 \times 10^{-13}$ erg cm$^{-2}$ s$^{-1}$, with filled squares
showing the values at selected fluxes to show the errors involved. In both
cases these define complete, X-ray flux limited LBL samples. The solid line
represents the X-ray number counts for LBL predicted by \cite{gio94} and
revised by \cite{pad95a}, while the dashed line shows the predictions of the
blazar sequence \citep{fos97}. See text for details.\label{blxcounts}}
\end{figure}

To see what happens in the X-ray band we have defined an X-ray flux limited
sample of LBL by applying a cut to our DXRBS sample at $f_{\rm x} \sim 1.6
\times 10^{-13}$ erg cm$^{-2}$ s$^{-1}$ (dot-dashed line in
Fig. \ref{fxfr}, bottom).  The resulting 11 objects form a complete, X-ray
selected LBL sample which reaches approximately the same limiting flux as
the EMSS sample. The X-ray number counts for this sample are shown in
Fig. \ref{blxcounts} (dotted line, filled squares) to be in extremely good
agreement with the predictions of \cite{pad95a} (solid line)\footnote{These
number counts predictions were originally published by \cite{gio94} and
then revised, but not published in this form, by \cite{pad95a}.}. Note also
how the LBL/HBL ratio increases roughly sevenfold at fainter fluxes, going
from $\sim 1 - 2 \%$ at $f_{\rm x} \sim 10^{-12} - 10^{-11}$ erg cm$^{-2}$
s$^{-1}$ to $\sim 10\%$ at $f_{\rm x} \sim 10^{-13}$ erg cm$^{-2}$
s$^{-1}$. For comparison, \cite{fos97} predicted the opposite behavior,
that is a slight ($5 - 10 \%$) {\it decrease} over the same flux range
(dashed line in Fig. \ref{blxcounts}). We also evaluated the X-ray LF for
the DRXBS LBL sample and, again, found it to be in very good agreement with
the predictions made by \cite{pad95a} (see their Fig. 4).

All available evidence seems then to favor a scenario where HBL represent a
small ($\approx 10\%$), constant fraction of the BL Lac population,
contrary to the predictions of the blazar sequence. This adds to the recent
questioning of the general validity of the blazar sequence
\citep[e.g.,][see also \cite{pad07}]{gio99,pad03,cac04,ab05,nie06,lan06}.
Note that in this case, unlike that for FSRQ and BL Lacs (\S~\ref{blfs_sec}), 
relative number counts do translate into relative space densities, as none
of the two classes evolve. 

\subsection{FSRQ}\label{fscounts_sect}

%fig. 6
\begin{figure}%6
%\plotone{f6.eps}
\centerline{\includegraphics[width=9.5cm]{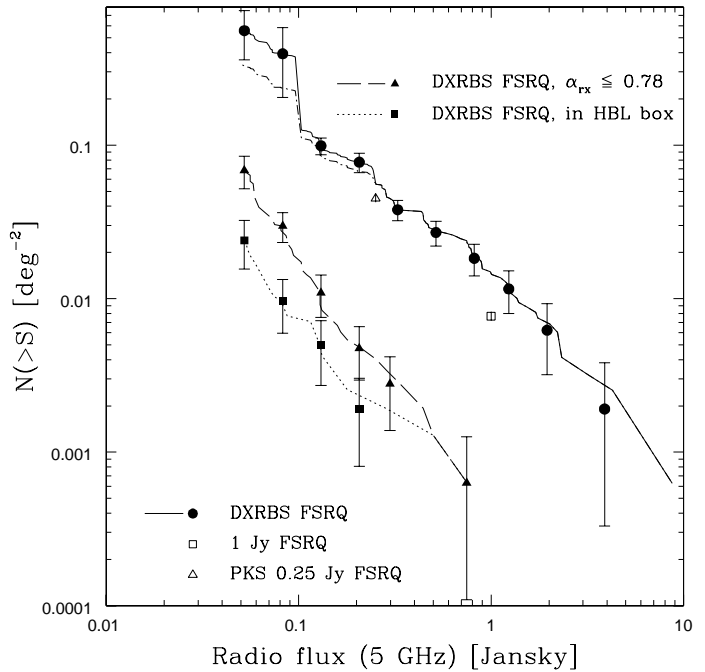}}
\caption{The integral number counts for DXRBS FSRQ at 5 GHz (6 cm). The
solid line represents the total number counts, while filled circles show
the values at selected fluxes to show the errors involved. The
dotted-dashed line shows the counts without the correction which we applied
to make DXRBS equivalent to a radio flux-limited sample
(\S~\ref{xrayradio}).  The open square represents the surface density for
the 1 Jy sample while the open triangle represents the surface density for
the PKS 0.25 Jy sample. The dashed line and filled triangles are the number
counts for the FSRQ with $\alpha_{\rm rx} \le 0.78$, while the dotted line
and filled squares are the number counts for the FSRQ in the HBL box. See
text for details.\label{fscounts}}
\end{figure}

Fig. \ref{fscounts} presents the integral number counts at 5 GHz for the
DXRBS FSRQ (solid line), compared to the values derived from the 1 Jy
\citep[][open square]{sti94} and the PKS 0.25 Jy \citep[][open triangle]
{wal05} samples. The latter value has been derived by converting the
numbers given in \cite{wal05} to 5 Ghz assuming $\alpha_{\rm r} \sim 0$ and
by multiplying them by the ratio of DXRBS FSRQ with $\alpha_{\rm r} \le
0.5$ and those with $\alpha_{\rm r} \le 0.4$ ($\sim 1.2$), to compensate
for the fact that the \cite{wal05} sample includes only sources with
$\alpha_{\rm r} \le 0.4$. As before, the filled circles show the values at
selected bins to show the errors involved without crowding the plot.

The following points can be made regarding Fig. \ref{fscounts}: 1. our
number counts agree with previous estimates at 250 mJy; they appear to be
higher than previous surveys at 1Jy but, given the error bars, not
significantly so (although this might be related to the effect discussed in
\S~\ref{serend}); 2. the number ratio between FSRQ and BL Lacs appear to be
independent of radio flux and $\approx 8$ (taking into account the possible
excess of DRXBS FSRQ at high fluxes). As for BL Lacs, the southern and
northern counts agree within the errors down to $\sim 150 - 200$ mJy, where
the effect of the higher flux limit of the northern hemisphere starts to
become relevant.

We have also derived number counts for "HBL-like" FSRQ, or HFSRQ, defined
as for BL Lacs in terms of their $\alpha_{\rm rx}$ and their position in
the $\alpha_{\rm ox}$, $\alpha_{\rm ro}$ plane. These are shown in
Fig. \ref{fscounts}. The fraction of FSRQ with $\alpha_{\rm rx} \le 0.78$
is $12^{+3}_{-2}\%$, while that in the HBL box is is $4^{+2}_{-1}\%$. The
fraction of HFSRQ, according to both definitions, appears also to be
roughly constant with radio flux, within the errors, as was the case for
HBL.

We note that our derived number counts have important implications also for
the study of the Cosmic Microwave Background (CMB), as DXRBS reaches radio
fluxes which are much fainter than those of the foreground sources detected
in the WMAP catalogue. \cite{gio06} have indeed shown that our surface
densities imply that a significant number of faint blazars are expected to
contaminate CMB fluctuation maps as foreground sources, with important
implications for the CMB fluctuation spectrum.

\section{Evolution}\label{evolution}
\begin{deluxetable*}{lrlcccc}
\tabletypesize{\scriptsize}
\tablecaption{DXRBS Evolutionary Properties. \label{tab2}}
\tablewidth{0pt}
\tablehead{
 ~&~&~&\multispan2{$H_0 = 70$, $\Omega_{\rm M} = 0.3$, $\Omega_{\rm
 \Lambda} = 0.7$~~~~}& \multispan2{~~~~$H_0 = 50$, $\Omega_{\rm M} = 0$,
 $\Omega_{\rm \Lambda} = 0$}\\
\colhead{Sample}&\colhead{N}&\colhead{$\langle z
\rangle$}&\colhead{$\langle V_{\rm e}/V_{\rm
a}\rangle$}&\colhead{$\tau$}&\colhead{$\langle V_{\rm e}/V_{\rm
a}\rangle$}&\colhead{$\tau$}\\}
\startdata
FSRQ & 129 & $1.82\pm0.08$ & $0.635\pm0.025$ & $0.27^{+0.05}_{-0.03}$
&$0.621\pm0.025$ & $0.26^{+0.05}_{-0.04}$\\
FSRQ, $f_{6 {\rm cm}} \ge 100$ mJy & 102 & $1.40\pm0.07$ & $0.608\pm0.029$ &
$0.33^{+0.09}_{-0.06}$ &$0.593\pm0.029$ & $0.32^{+0.13}_{-0.06}$\\
FSRQ, $f_{\rm x}/f_{\rm r} \ge 4 \times 10^{-13}$ & 101 & $1.55\pm0.07$ &
$0.724\pm0.029$ & $0.18^{+0.02}_{-0.02}$ &$0.723\pm0.029$ &
$0.16^{+0.02}_{-0.01}$\\
FSRQ, $z \le 1$ & 49 & $0.71\pm0.03$ & $0.629\pm0.041$ &
$0.18^{+0.07}_{-0.04}$ &$0.640\pm0.041$ & $0.14^{+0.05}_{-0.03}$\\
FSRQ, $\alpha_{\rm rx} \le 0.78$ (HFSRQ) & 37 & $1.79\pm0.16$ &
$0.735\pm0.047$ & $0.20^{+0.03}_{-0.03}$&$0.726\pm0.047$ &
$0.17^{+0.03}_{-0.02}$\\
FSRQ, in HBL box & 15 & $1.59\pm0.23$ & $0.740\pm0.075$ &
$0.19^{+0.05}_{-0.03}$&$0.724\pm0.075$ & $0.16^{+0.05}_{-0.03}$\\
BL Lacs & 24 & $0.26\pm0.04$\tablenotemark{a} &
$0.54\pm0.06$\tablenotemark{b} & ...\tablenotemark{c} &
$0.57\pm0.06$\tablenotemark{b} & ...\tablenotemark{c} \\
BL Lacs, $f_{6 {\rm cm}} \ge 100$ mJy & 17 & $0.29\pm0.05$\tablenotemark{d}
& $0.42\pm0.07$\tablenotemark{b} & ...\tablenotemark{c} &
$0.45\pm0.07$\tablenotemark{b} & ...\tablenotemark{c} \\
BL Lacs, $f_{\rm x}/f_{\rm r} \ge 4 \times 10^{-13}$ & 22 &
$0.26\pm0.04$\tablenotemark{d} & $0.54\pm0.06$\tablenotemark{b} &
...\tablenotemark{c} & $0.56\pm0.06$\tablenotemark{b} &
...\tablenotemark{c} \\
BL Lacs, $\alpha_{\rm rx} \le 0.78$ (HBL) & 7 &
$0.35\pm0.05$\tablenotemark{e} & $0.54\pm0.10$\tablenotemark{b} &
...\tablenotemark{c} & $0.56\pm0.10$\tablenotemark{b} &
...\tablenotemark{c} \\
BL Lacs, $\alpha_{\rm rx} > 0.78$ (LBL) & 17 &
$0.19\pm0.03$\tablenotemark{f} & $0.54\pm0.07$\tablenotemark{b} &
...\tablenotemark{c} & $0.57\pm0.07$\tablenotemark{b} &
...\tablenotemark{c} \\
BL Lacs, in HBL box & 6 & $0.37\pm0.06$\tablenotemark{g} &
$0.48\pm0.13$\tablenotemark{b} & ...\tablenotemark{c} &
$0.50\pm0.13$\tablenotemark{b} & ...\tablenotemark{c} \\
\enddata
\tablenotetext{a}{Excluding the 7 sources without redshift}
\tablenotetext{b}{Assuming $z = \langle z \rangle$ for the sources 
without redshift}
\tablenotetext{c}{$\langle V_{\rm e}/V_{\rm a}\rangle$ not significantly 
different from 0.5: no evolution assumed}
\tablenotetext{d}{Excluding 6 sources without redshift}
\tablenotetext{e}{Excluding 2 sources without redshift}
\tablenotetext{f}{Excluding 5 sources without redshift}
\tablenotetext{g}{Excluding 1 source without redshift}
\end{deluxetable*}

We study the evolutionary properties of DXRBS through the 
$V_{\rm e}/V_{\rm a}$ test \citep{av80, mor91}, a variation of 
the $V/V_{\rm max}$ test \citep{sch68}, that is the ratio between 
{\it enclosed} and {\it available} volume. 
Values of $\langle V_{\rm e}/V_{\rm a} \rangle$
significantly different from 0.5 indicate evolution, which will be positive
(i.e., sources were more luminous and/or more numerous in the past) for
values $> 0.5$, and negative (i.e., sources were less luminous and/or less
numerous in the past) for values $< 0.5$. Moreover, one can also fit an
evolutionary model to the sample by finding the evolutionary parameter
which makes $\langle V_{\rm e}/V_{\rm a} \rangle = 0.5$.

We have computed $V_{\rm e}/V_{\rm a}$ values for our sources taking into
account our flux limits (at 6 and 20 cm and in the X-ray band) and the
appropriate sky coverage. Statistical errors are given by $\sigma =
1/\sqrt{12~N}$ \citep{av80}. To have a first, simple estimate of the sample
evolution we have also derived the best fit parameter $\tau$ assuming a
pure luminosity evolution of the type normally used, i.e., $P(z) = P(0){\rm
exp}[T(z)/\tau]$, where $T(z)$ is the look-back time (the smaller $\tau$
the stronger the evolution).

Table \ref{tab2} gives the sub-sample in
column (1), the number of sources in column (2), the mean redshift in
column (3), $\langle V_{\rm e}/V_{\rm a} \rangle$ and $\tau$ in columns (4)
and (5) for our $\Lambda$ cosmology, and $\langle V_{\rm e}/V_{\rm a}
\rangle$ and $\tau$ in columns (6) and (7) for the empty Universe cosmology,
for comparison with previous results. Note that $\langle z \rangle$ is
calculated taking into account the effect of the sky coverage (see
discussion in Paper II).

The main results are the following:
\begin{enumerate}
\item DXRBS FSRQ evolve at the $5.4\sigma$ ($4.8\sigma$ for an empty
Universe cosmology) level, a well known result \citep[e.g.,][]{pau92,up95}.
Their evolutionary parameter for the simple case of pure luminosity
evolution is fully consistent with that derived by \cite{up95} for the 2 Jy
FSRQ sample for an empty Universe ($\tau = 0.23^{+0.07}_{-0.04}$);
\item DXRBS BL~Lacs do not evolve, i.e., their $\langle V_{\rm e}/V_{\rm a}
\rangle$ value is not significantly different from 0.5 (and consequently
$\tau \ga 1$). The results for BL~Lacs are more uncertain because of the
smaller number statistics. The fact that $\sim 30$\% of them have no
redshift is less of a problem, as redshift affects $V_{\rm e}/V_{\rm a}$ 
values way less than flux (a value equal to $\langle z \rangle$ was 
assumed in this case);
\item The typical redshifts for FSRQ and BL Lacs are markedly different,
with the former covering the $0.2 - 4.7$ range and having $\langle z
\rangle = 1.82\pm0.08$ and the latter having $0.04 \le z \le 0.73$ and
$\langle z \rangle = 0.26\pm0.04$;
\item The $\langle V_{\rm e}/V_{\rm a} \rangle$ values for HBL and LBL
(using both definitions; see \S~\ref{counts}) are not significantly
different. This is a new result, which contradicts the commonly accepted
point of view that HBL and LBL have different evolutionary properties.
Notice that {\it for the first time} we can study the evolution of HBL and
LBL {\it within the same sample}. Previous comparisons had been typically
made between the 1~Jy (radio-selected) and the EMSS samples
(X-ray-selected), although \cite{rec00} did study the dependency of
$\langle V_{\rm e}/V_{\rm a} \rangle$ on X-ray-to-radio flux ratios for the
EMSS sample. Admittedly, the errors on the $\langle V_{\rm e}/V_{\rm a}
\rangle$ values are rather large but this is the best that can be done at
present;
\item HFSRQ have mean redshifts and evolutions similar to those of the main
FSRQ sample. Although the $\langle V_{\rm e}/V_{\rm a} \rangle$ values
appear slightly larger for HFSRQ (but still within $\la 2\sigma$), this
difference can be easily explained by the fact that the HFSRQ sample is
free from the effect discussed in \S~\ref{xrayradio}, unlike the full
sample (see below), and completely identified, as all still to be observed
sources have $\alpha_{\rm rx} > 0.78$ and are outside the HBL box.

\end{enumerate} 

To check for the effect of the corner of the radio flux -- X-ray flux plane
"missed" by DXRBS (\S~\ref{xrayradio}), which is necessary
only if one wants to compare our results to those of purely radio
flux-limited samples, one cannot take the approach we used in
\S~\ref{counts}. The $V_{\rm e}/V_{\rm a}$ test, in fact, involves
redshift, and so a simple correction to the observed number of sources with
a given radio flux will not be sufficient. We have taken two complementary
approaches to address this: 1. we evaluated $\langle V_{\rm e}/V_{\rm a}
\rangle$ for both FSRQ and BL Lac samples applying a cut in radio flux at
100 mJy. By doing this the corner that DXRBS is missing in the radio--X-ray
flux plane in Fig. \ref{fxfr} shrinks considerably, although we do lose a
factor of two in radio flux depth. Tab. \ref{tab2} shows that $\langle
V_{\rm e}/V_{\rm a} \rangle$ for these higher radio flux samples is
consistent with that for the full samples within $\sim 1.3\sigma$; 2. we
defined a sub-sample with complete coverage of the radio--X-ray flux plane
for $f_{\rm x}/f_{\rm r} \ge 4 \times 10^{-13}$ erg cm$^{-2}$ s$^{-1}$
Jy$^{-1}$ (to the right of the dotted line in Fig. \ref{fxfr}, bottom; this
translates into an additional constraint on the X-ray flux limit, which is
bound to be $\ge 4 \times 10^{-13} \times (f_{\rm r}/{\rm Jy})$ erg
cm$^{-2}$ s$^{-1}$). By doing this DXRBS has {\it complete} coverage of a
somewhat {\it restricted} region of the radio--X-ray flux plane down to our
nominal radio flux limit.  Tab. \ref{tab2} shows that $\langle V_{\rm
e}/V_{\rm a} \rangle$ for these sub-samples is slightly larger than for the
full samples but still within $\sim 2 \sigma$ for our adopted
cosmology. Both these results show that the double X-ray/radio selection
discussed in \S~\ref{xrayradio} has only a small effect on the derivation
of the evolutionary properties of DXRBS.  This makes sense as our X-ray
flux limit, which is above the one which would guarantee complete coverage
of the radio--X-ray flux plane, is properly taken into account when
calculating the volumes (but this was not the case when deriving number
counts).

We have also assessed the effect of the still unidentified sources as
follows. We have assumed that all remaining 10 sources with $\alpha_{\rm
r} \le 0.5$ are FSRQ and all remaining 6 sources with $\alpha_{\rm r}
> 0.5$ are BL Lacs. (Note that based on the relative fraction of the
different DXRBS classes the expected numbers would be 9 and 2 for FSRQ and
BL Lacs respectively: see \S~\ref{definition}). We then assigned a redshift
equal to $\langle z \rangle$ and then added these sources to the FSRQ and
BL Lac samples. The resulting $\langle V_{\rm e}/V_{\rm a} \rangle$ values
increase only by $\la 0.6\sigma$ as compared to the values given in
Tab. \ref{tab2}, which shows that our results are quite stable against the
addition of the still unidentified objects.

Given the available FSRQ statistics we can move beyond the assumption of
pure luminosity evolution and study in detail possible redshift
dependencies. In the case of a pure (luminosity or density) evolution
model, that is under the assumption that the rate of change is independent
of cosmic epoch, the best fit evolutionary parameter has to be the same at
all redshifts \citep[see, e.g.,][]{de92}. It then follows that $\langle
V_{\rm e}/V_{\rm a} \rangle$ has also to be constant. The simplest test we
can do is then to see if, and how, $\langle V_{\rm e}/V_{\rm a} \rangle$
changes with redshift. We then split our FSRQ sample in six redshift bins
so that each bin contained roughly the same number of sources (22 for the
first three bins and 21 for the last three). $\langle V_{\rm e}/V_{\rm a}
\rangle$ is roughly constant and $\sim 0.6$ between $z \sim 0.3$ and 2,
while above this redshift it drops to 0.47 (from 0.65 at $z \sim 0.3$), a
change which is significant at the $2\sigma$ level. The sign of the
evolution also changes, from strongly ($2.5\sigma$) positive at $z \sim
0.3$ to consistent with no evolution at $z \ga 2$.

%fig. 7
\begin{figure}%7
%\plotone{f7.eps}
\centerline{\includegraphics[width=9.5cm]{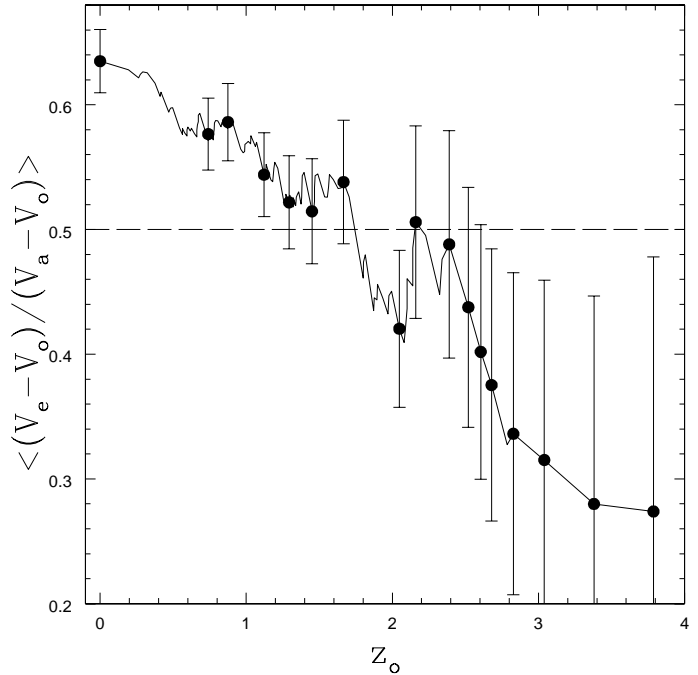}}
\caption{The banded $\langle V_{\rm e}/V_{\rm a} \rangle$ statistic,
$\langle (V_{\rm e} - V_o)/(V_{\rm a} - V_o) \rangle$ versus $z_o$ for
DXRBS FSRQ. The horizontal dashed line indicates the value of 0.5 expected
under the null hypothesis of no evolution. The ($1~\sigma$) statistical
error bars, given by $\sigma = 1/\sqrt{12~N}$, where $N$ is the number of
objects in the subsample with $z > z_o$, are shown for selected
redshifts. \label{vevabanded}}
\end{figure}

An alternative, more common way to check for a cosmic epoch-dependent
evolution is through the so-called banded $\langle V_{\rm e}/V_{\rm a}
\rangle$ statistic, i.e., $\langle (V_{\rm e} - V_o)/(V_{\rm a} - V_o)
\rangle$, where $V_o$ is the cosmological volume enclosed by a redshift
$z_o$ \citep[see, e.g.,][]{du90}. This allows the detection of any
high-redshift, possibly negative evolution by separating it by the
well-known strong, positive, low-redshift evolution. This is shown in Fig.
\ref{vevabanded}. The change is very strong, with a highly significant drop
in $\langle V_{\rm e}/V_{\rm a} \rangle$ with redshift, and evolution
vanishing by $z \ga 1.3$. $\langle V_{\rm e}/V_{\rm a} \rangle$, in fact,
goes from being $> 0.5$ at the $5.4\sigma$ level at $z \sim 0$ to values
which are $< 0.5$, although not significantly so, for $z \ga 2$. At higher
redshifts, in fact, the lack of sources makes it very difficult to accrue
more meaningful statistics from this test. This is an inherent limitation
of DXRBS, due to the fact that the faintest X-ray sources are only detected
in a relatively small area of the sky (see Fig. \ref{skycov}). Since, on
average, these are also the sources with the highest redshifts, this
explains our relatively large error bars at $z \ga 2$.
These results, however, are consistent, for example, with
those of \cite{du90}, \cite{jar00}, and \cite{ars06}. Stronger evidence for
a redshift cut-off comes from the evolution of the FSRQ luminosity function
(\S~\ref{fslf}).

As regards BL Lacs, due to their smaller number, we did two simpler checks:
first, we divided the sample with redshift information into two bins
containing an equal number of sources, above and below $z = 0.264$; the
$\langle V_{\rm e}/V_{\rm a} \rangle$ values for the two samples are both
consistent with 0.5 within $2\sigma$; second, we applied the banded
$\langle V_{\rm e}/V_{\rm a} \rangle$ statistics for $z_o = 0$ (whole
sample) and $z_o = 0.264$; again, the two values are consistent with 0.5,
this time within $1.3\sigma$. In the following we then assume no evolution
at all redshifts for the BL Lac sample.

\section{Luminosity Functions}\label{lf}

As for the case of the evolutionary properties, the double X-ray/radio
selection discussed in \S~\ref{xrayradio} has only a very small effect on
the derivation of the DXRBS LFs. In particular, by adopting the two
approaches discussed in \S~\ref{evolution} (i.e., applying a radio cut at 
100 mJy and defining a sub-sample with $f_{\rm x}/f_{\rm r} \ge 
4 \times 10^{-13}$ erg cm$^{-2}$ s$^{-1}$) there is little, if any, change
in the BL Lac and FSRQ LFs, typically well within $1\sigma$ at a given
radio power. This is because the "missing" low radio flux sources would be
spread over a range of powers, making the overall impact smaller. Moreover,
the volumes are properly calculated using our X-ray flux limit (see
\S~\ref{evolution}). All our LFs are derived using the $1/V_{\rm max}$ (in our
case $1/V_{\rm a}$) technique \citep{sch68}. 

\subsection{BL Lacs}\label{bllf}

%fig. 8
\begin{figure}%8
\centerline{\includegraphics[width=9.5cm]{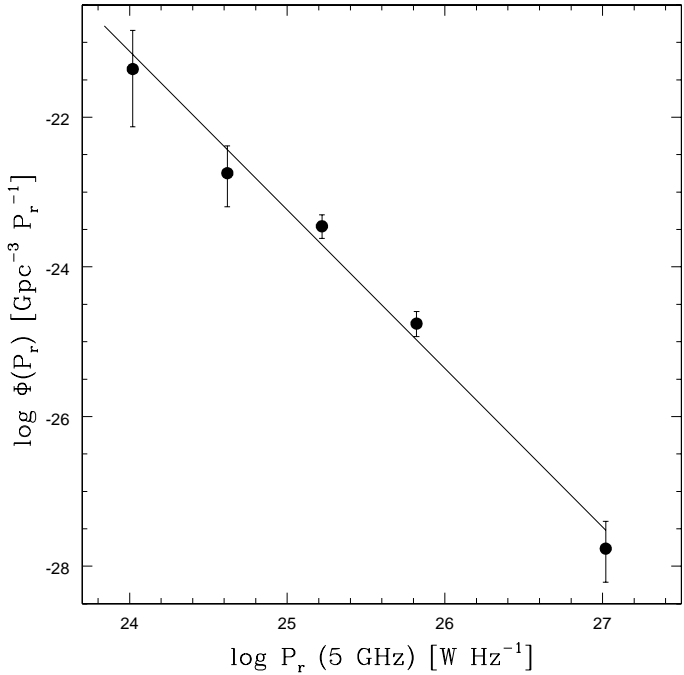}}
%\plotone{f8.eps}
\caption{The radio luminosity function of DXRBS BL Lacs (filled
points). Error bars correspond to $1\sigma$ Poisson errors
\citep{geh86}. The solid line is a weighted least-squares fit to the
data. \label{lfbl}}
\end{figure}
The LF of DXRBS BL Lacs is shown in Fig. \ref{lfbl}. As discussed above,
based on the $\langle V_{\rm e}/V_{\rm a} \rangle$ analysis, no evolution
is assumed. Therefore, the shown LF is supposed to be epoch-independent. We
have assumed $z = \langle z \rangle \sim 0.26$ for the 7/24 BL Lacs without
redshift. 

The LF is well fitted by a single power law of the form $\phi(P_{\rm r})
\propto P_{\rm r}^{-B_r}$. Varying the binning, the differential slope is
in the range $2.05 < B_r < 2.23$. For a bin size $\Delta \log P = 0.6$,
which is representative, a weighted least-squares fit yields $\phi(P_{\rm
r}) \propto P_{\rm r}^{-2.12\pm0.16}$ ($\chi^2_{\nu} \sim 1.2$ for 3
degrees of freedom). The total number density of BL Lacs in the range $7
\times 10^{23} - 6 \times 10^{26}$ W/Hz, derived independently of bin size
from the integral LF, is $840\pm100$ Gpc$^{-3}$.

We have checked how our assumption on the missing redshifts affects the LF
determination in two ways: 1. we have excluded the 7 objects without
redshift from our computation and multiplied the LF by 24/17. The resulting
LF is consistent with the previous one within the errors, which shows that
the overall shape of the LF is not strongly dependent on us assuming $z =  
\langle z \rangle$ for the 7 BL Lacs without redshift;
2. the fact that these sources show a featureless continuum might suggest
that their redshifts could be the highest in the sample.  The lack of
features, in fact, could be related to stronger beaming and, therefore, to
(on average) higher luminosities. We have then obtained lower limits on the
redshifts of these sources from their V magnitudes by making use of the
fact that BL Lacs are hosted by ellipticals of almost constant luminosity
\citep[e.g.,][]{ur00}, which makes them adequate standard candles. By
assuming a value for the ratio of the jet/galaxy flux when a BL Lac appears
featureless, one can estimate the apparent magnitude of the host galaxy,
which gives in turn a lower limit on the redshift of the BL Lac (as the
jet/galaxy ratio could be higher).  We have taken a conservative limit on
the jet/galaxy ratio of a featureless BL Lac of one, based on Fig. 1 of
\cite{lan02}, and have used the relation between apparent V magnitude and
redshift for luminous ellipticals published by \cite{bro93}. The resulting
lower limits span the range $0.11 - 0.64$, with $\langle z \rangle \sim
0.28$, very close to the value we assumed. With a jet/galaxy ratio of 10,
which based on \cite{lan02} could be more appropriate for a featureless BL
Lac, we obtain lower limits in the range $0.24 - 1.47$, with $\langle z
\rangle \sim 0.65$. However, even assuming this value for the missing
redshifts changes very little the resulting LF, which is consistent with the
previous one well within the errors.

%fig. 9
\begin{figure}%9
%\plotone{f9.eps}
\centerline{\includegraphics[width=9.5cm]{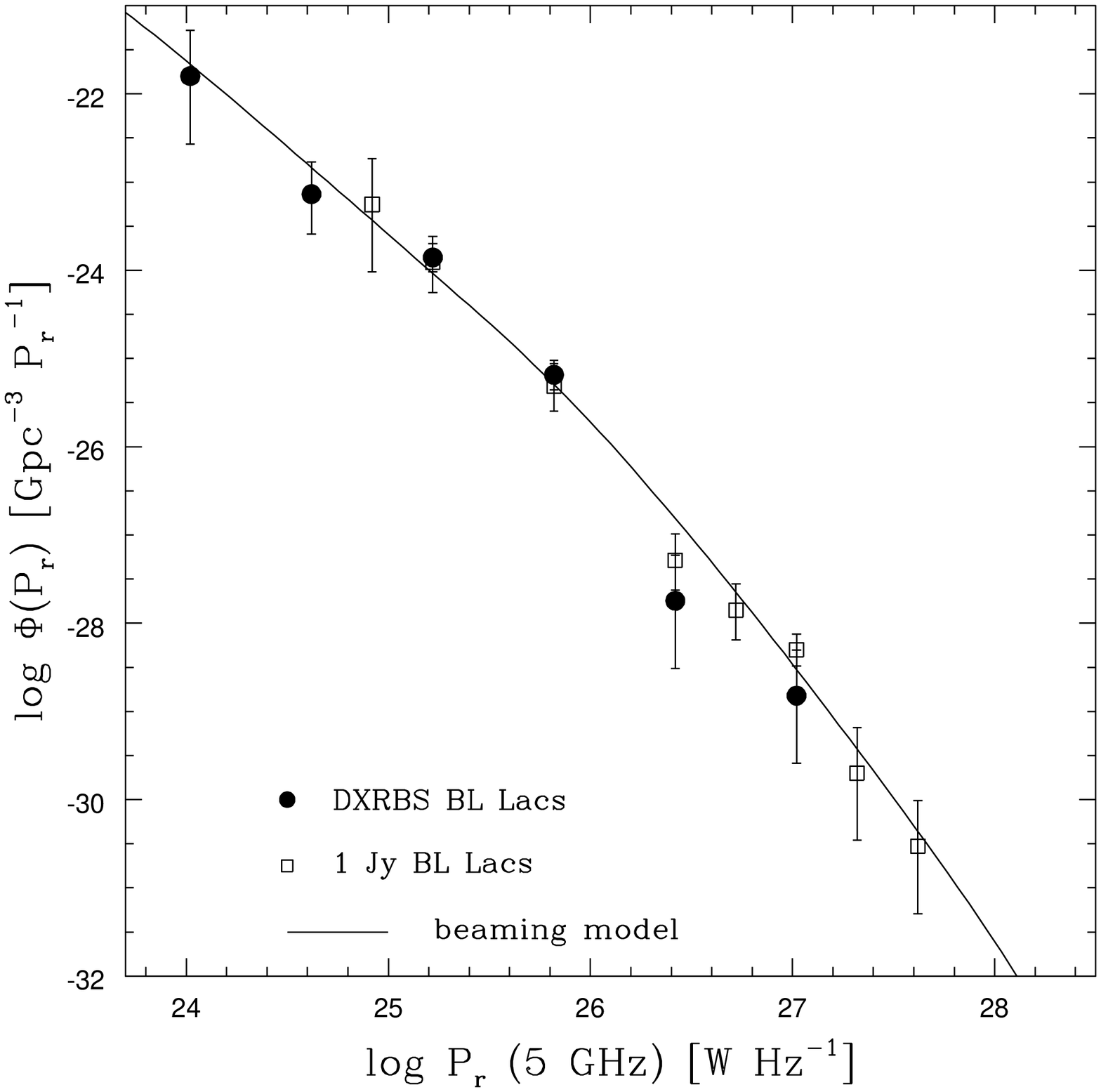}}
\caption{The radio luminosity function of DXRBS BL Lacs (filled points)
compared to the predictions of a beaming model based on the 1 Jy luminosity
function and evolution \citep[solid line,][]{up95}. The open squares
represent the 1 Jy luminosity function \citep{sti91}. Error bars correspond
to $1\sigma$ Poisson errors \citep{geh86}. The beaming predictions and the
1 Jy points have been corrected to take into account our somewhat different
definition of a BL Lac. For consistency with \citet{sti91} the DXRBS
luminosity function has been de-evolved to zero redshift and an $H_0 = 50$
km s$^{-1}$ Mpc$^{-1}$, $\Omega_{\rm M} = 0$, and $\Omega_{\rm \Lambda} =
0$ cosmology has been adopted. See text for details.\label{lfblbeaming}}
\end{figure}

Fig. \ref{lfblbeaming} shows the LF of DXRBS BL Lacs for an empty Universe
cosmology (filled points), to compare it with previous determinations and
the predictions of unified schemes. The figure shows also the 1 Jy LF
\citep[open squares,][]{sti91}, and the predictions of a beaming model
based on the 1 Jy LF and evolution \citep[solid line,][]{up95}. These show
what one should expect to find when reaching powers lower than those used
to constrain the LF at the high end. Note that since \cite{sti91} had
de-evolved their LF to zero redshift using their best fit value ($\tau =
0.32$), we have done the same for DXRBS BL Lacs, for which we get $\tau =
0.45^{+3.05}_{-0.18}$.

A few interesting points can be made: 1. the  DXRBS and 1 Jy LFs are in very
good agreement in the region of overlap, despite the factor $\sim 20$
difference in limiting flux; 2. the DXRBS LF reaches powers about one order
of magnitude smaller that those reached by the 1 Jy LF, as expected given
point n. 1; 3. the DXRBS LF is in good agreement with the predictions of
unified schemes, which means that the unification of BL Lacs and
Fanaroff-Riley (FR) type I radio galaxies seems to work also at low powers.
Note that while the 1 Jy LF was fitted by $\phi(P_{\rm r}) \propto P_{\rm
r}^{-2.53\pm0.15}$ \citep{sti91}, the DXRBS LF is somewhat flatter, with
$\phi(P_{\rm r}) \propto P_{\rm r}^{-2.31\pm0.18}$ (for a bin size $\Delta
\log P = 0.6$; $\chi^2_{\nu} \sim 2.7$ for 4 degrees of freedom; as before,
the slope does not change, well within the errors, for different bin
sizes). This is explained by the fact that the BL Lac LF predicted by
unified schemes flattens out for $P_{\rm r} \la 10^{26}$ W/Hz, a region
which is better sampled by DXRBS. For this cosmology the total number
density of BL Lacs in the range $10^{24} - 6 \times 10^{26}$ W/Hz, derived
independently of bin size from the integral LF, is $310\pm35$ Gpc$^{-3}$,
to be compared with the value of $40$ Gpc$^{-3}$ in the range $6 \times
10^{24} - 3 \times 10^{27}$ W/Hz for the 1 Jy LF.

Two caveats are worth mentioning: 1. as done above, we have assumed $z =
\langle z \rangle \sim 0.26$ for the 7/24 BL Lacs without redshift but, as
before, the overall shape of the LF is not strongly dependent on this
assumption; 2. as mentioned above (\S~\ref{counts}) the definition of a BL
Lac for the 1 Jy and DXRBS samples is somewhat different. Since $2/3$
(16/24) of the DXRBS BL Lacs fulfill the 1 Jy definition, the 1 Jy points
and the beaming predictions based on them were multiplied by the factor
$3/2$ to take this difference into account.

\subsection{FSRQ}\label{fslf}

%fig. 10
\begin{figure}%10
%\plotone{f10.eps}
\centerline{\includegraphics[width=9.5cm]{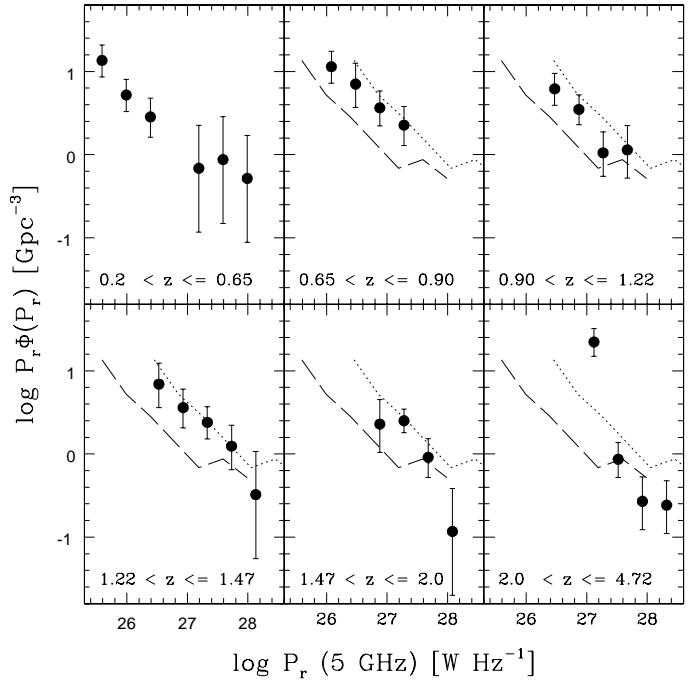}}
\caption{The differential radio luminosity function of DXRBS FSRQ in a $P
\times \phi(P)$ form in six different redshift bins (including roughly the
same number of sources): $0.2 - 0.65$, $0.65 - 0.90$, $0.90 - 1.22$, $1.22
- 1.47$, $1.47 - 2.0$, and $2.0 - 4.72$. The dashed line indicates, for
reference, the luminosity function in the lowest redshift bin, while the
dotted line shows what the highest redshift LF should be in the case of
pure luminosity evolution with $\tau$ equal to the best fit parameter given
in Tab. \ref{tab2} for the whole FSRQ sample. Error bars correspond to
$1\sigma$ Poisson errors \citep{geh86}. See text for details.\label{lffs}}
\end{figure}

%fig. 11
\begin{figure}%11
%\plotone{f11.eps}
\centerline{\includegraphics[width=9.5cm]{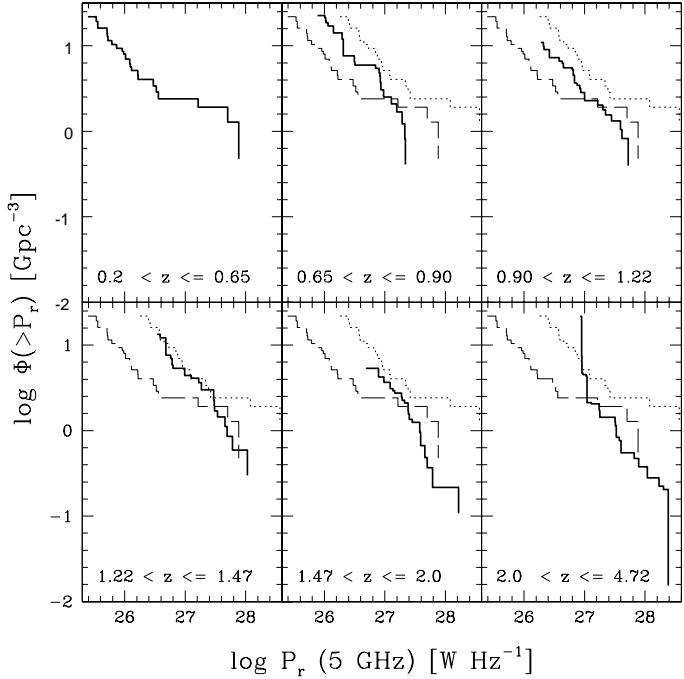}}
\caption{The integral radio luminosity function of DXRBS FSRQ in six
different redshift bins (including roughly the same number of sources):
$0.2 - 0.65$, $0.65 - 0.90$, $0.90 - 1.22$, $1.22 - 1.47$, $1.47 - 2.0$,
and $2.0 - 4.72$. The dashed line indicates, for reference, the luminosity
function in the lowest redshift bin, while the dotted line shows what the
highest redshift LF should be in the case of pure luminosity evolution with
$\tau$ equal to the best fit parameter given in Tab. \ref{tab2} for the
whole FSRQ sample. See text for details.\label{intlffs}}
\end{figure}

The case for FSRQ is more complex, as we know from the $\langle (V_{\rm e}
- V_o)/(V_{\rm a} - V_o) \rangle$ analysis (\S~\ref{evolution}) that the
evolutionary parameter is epoch-dependent, and therefore we cannot simply
de-evolve the global LF to zero redshift. We have then studied the LF
evolution as a function of redshift. To be able to do so in a meaningful
way retaining also a significant number of sources per redshift bin we have
divided our sample in six bins so that each bin contains roughly the same
number of sources (22 for the first three bins and 21 for the last three).
Finally, we have computed both the differential and integral LFs, as they
give complementary information. 

%fig. 12
\begin{figure}%12
%\plotone{f12.eps}
\centerline{\includegraphics[width=9.5cm]{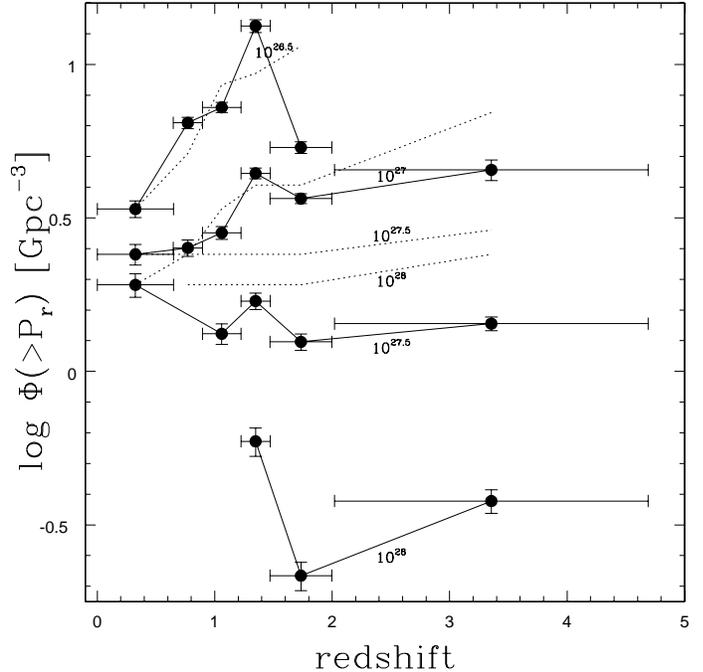}}
\caption{Integral number densities for DXRBS FSRQ as a function of redshift
for four power ranges, from top to bottom: $P_{\rm r} > 10^{26.5}$ W/Hz,
$P_{\rm r} > 10^{27}$ W/Hz, $P_{\rm r} > 10^{27.5}$ W/Hz, $P_{\rm r} >
10^{28}$ W/Hz. The dotted lines show the integral number densities expected
in the case of pure luminosity evolution with $\tau$ equal to the
best fit parameter given in Tab. \ref{tab2} for the whole FSRQ sample.
\label{lfintz}}
\end{figure}

Fig. \ref{lffs} shows the differential LF for DXRBS FSRQ in a $P \times
\phi(P)$ form. This is equivalent to the $\phi(M_{\rm B})$ form normally
used in the optical band and allows an easy separation of luminosity and
density evolution as the former would simply translate the LF to the right
(higher powers) with no change in the ordinate (number), while the opposite
would be true for the latter. The figure also shows (dotted line) what the
highest redshift LF should be in the case of pure luminosity evolution with
$\tau$ equal to the best fit parameter given in Tab. \ref{tab2} for the
whole FSRQ sample. Keeping in mind that for a single power law LF one
cannot distinguish between luminosity and density evolution, we assume that
some luminosity evolution takes place, based on studies in other bands
\citep[see, e.g.,][]{cro04}. Despite our somewhat limited statistics, and
related relatively large scatter, which prevent us from a more quantitative
analysis, three things are apparent: 1. most of the luminosity evolution
happens at relatively low redshift, as already by $z \sim 1.3$ the increase
in power is almost as high as expected in the highest redshift bin for a
simple pure luminosity evolution model. Beyond $z \approx 1$ not much
action seems to take place; 2. all but one of the six LFs can be fitted by
a single power-law with slope in the range 1.6 -- 1.9 ($\phi(P_{\rm r})
\propto P_{\rm r}^{-B_r}$; 4/6 are actually in the 1.6 -- 1.75 range), the
exception being the highest redshift one, whose anomalous steepness is due
to a single object. No trend of slope with redshift is present, within the
errors; 3. the total number of sources (maximum space density) appears to
be roughly constant at low redshifts ($z \la 1$).

Fig. \ref{intlffs} shows the integral LFs. Here the sharp shift in power
from $z \sim 0.4$ to $z \sim 1.3$ is clearer, as is the sudden stop at
higher redshifts. There is also evidence, for $P_{\rm r} \la 10^{27}$ W/Hz,
of an increase in the number density of FSRQ with redshift (at a given
power) and a hint of a decrease (with redshift) at larger powers. The
evolution of the number density is much better seen in Fig. \ref{lfintz},
which plots it as a function of redshift above four powers, labeled in the
figure. The curves show an initial, strong increase at low redshift/powers,
peaking at $z \approx 1.5$, followed by a decline at higher
redshifts. Moreover, while at low powers the integral number densities are
consistent with those expected in the case of pure luminosity evolution
(dotted lines), at high powers ($P \ga 3 \times 10^{27}$ W/Hz) they are
lower, pointing to a deficit of sources at high redshifts and luminosities
as compared to the pure luminosity evolution scenario.  All of the above
suggests a high-redshift ($z \approx 1.5$) decline in the comoving space
density of high-power ($P \ga 10^{27}$ W/Hz) FSRQ. These results are
similar to those obtained by \cite{wal05} for the Parkes 0.25 Jy sample.

We have also derived the LF of DXRBS FSRQ for an empty Universe cosmology,
to compare it with previous determinations of the local LF of FSRQ and the
predictions of unified schemes. Given that the evolutionary parameter is
epoch dependent (\S~\ref{evolution}) and that therefore we cannot simply
de-evolve the global LF to zero redshift, we have restricted ourselves to
sources having $z \le 1$. This is the redshift where the evolution appears
to slow down considerably (see Figs. \ref{lffs} and \ref{intlffs}) and it
gives us also a large enough number of objects. This sub-sample includes 49
sources and is characterized by $\tau = 0.14^{+0.05}_{-0.03}$ (see
Tab. \ref{tab2}). Note that this evolution is as strong as found in the
optical band for the 2dF QSO Redshift Survey (2QZ) \citep[$\tau = 1/k_1 =
1/6.15 \sim 0.16$;][]{cro04}.

%fig. 13
\begin{figure}%13
%\plotone{f13.eps}
\centerline{\includegraphics[width=9.5cm]{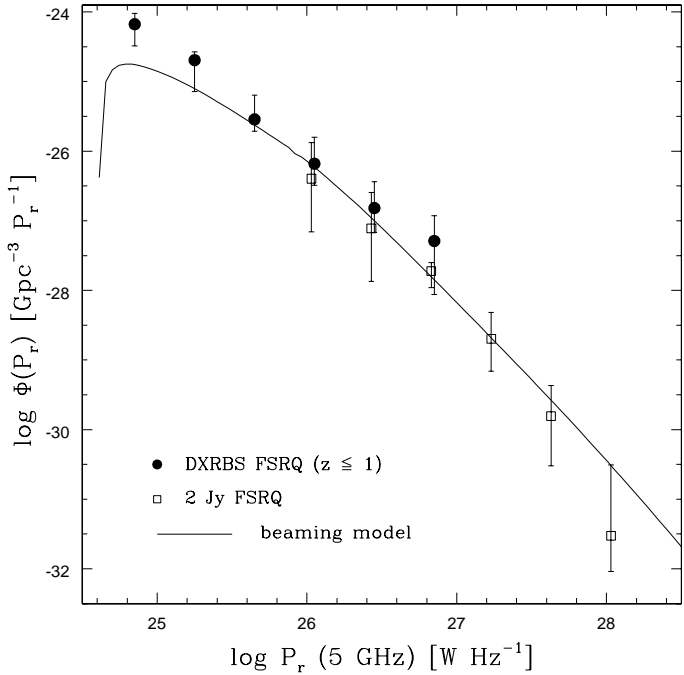}}
\caption{The radio luminosity function of DXRBS FSRQ with $z \le 1$ (filled
points) de-evolved at zero redshift compared to the predictions of a
beaming model based on the 2 Jy luminosity function and evolution
\citep[solid line,][]{up95}. The open squares represent the 2 Jy luminosity
function. Error bars represent the sum in quadrature of the $1\sigma$
Poisson errors \citep{geh86} and the variations of the number density
associated with a $1\sigma$ change in the evolutionary parameter
$\tau$. For consistency with \citet{up95} an $H_0 = 50$ km s$^{-1}$
Mpc$^{-1}$, $\Omega_{\rm M} = 0$, and $\Omega_{\rm \Lambda} = 0$ cosmology
has been adopted. Both beaming model and 2 Jy LF have been converted from
2.7 GHz assuming $\alpha_{\rm r} = 0$. See text for
details.\label{lffsbeaming}}
\end{figure}

The resulting LF is presented in Fig. \ref{lffsbeaming} (filled points),
which shows also the 2 Jy LF and the predictions of a beaming model based
on the 2 Jy LF and evolution \citep[solid line,][]{up95}. These show what
one should expect to find when reaching powers lower than those used to
constrain the LF at the high end. The model and 2 Jy LF have been 
converted from 2.7 GHz assuming $\alpha_{\rm r} = 0$. 

A few interesting points can be made: 1. the 2 Jy and DXRBS LFs are in very
good agreement in the region of overlap, despite the factor $\sim 40$
difference in limiting flux; 2. the DXRBS LF reaches powers more than one
order of magnitude smaller that those reached by the 2 Jy LF, as expected
given point n. 1; 3. the DXRBS LF is in good agreement with the predictions
of unified schemes of \cite{up95} , although slightly above them in the
first two luminosity bins, which means that the unification of FSRQ and FR
II radio galaxies seems to work also at low powers. The observed
differences could be simply due to the fact that the predictions are based
on a FR II LF derived from a very high flux (2 Jy) sample; 4. we are
getting close to the limits of the FSRQ "Universe". In fact, as FSRQ are
thought to be the beamed counterparts of high-power radio galaxies, the
low-luminosity part of their LF should end at relatively high
powers. Assuming that the minimum luminosity inferred from the fit to the 2
Jy LF is correct \citep[solid line in the figure, based on the 2 Jy LF of
FR II radio galaxies; see][]{up95}, then DXRBS is approaching that value.

Note that the model prediction of a flattening of the FSRQ LF for 
$P_{\rm r} \la 10^{26}$ W/Hz, a region which is better sampled by DXRBS, 
fits quite well the fact that the 2 Jy LF is $\phi(P_{\rm r}) \propto 
P_{\rm r}^{-2.3\pm0.3}$, while the DXRBS LF is flatter, with 
$\phi(P_{\rm r}) \propto P_{\rm r}^{-1.63\pm0.16}$. 

\subsection{BL Lacs and FSRQ}\label{blfs_sec}

\begin{figure}%14
%\plotone{f14.eps}
\centerline{\includegraphics[width=9.5cm]{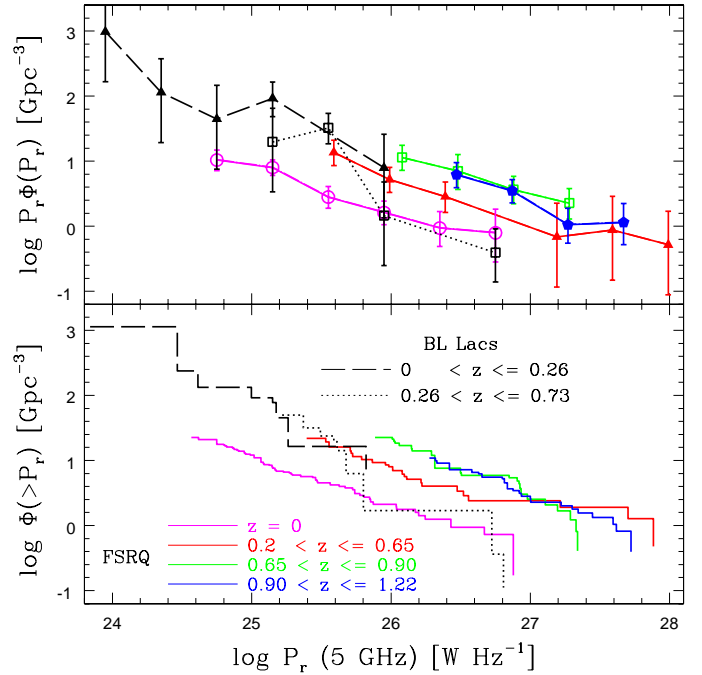}}
\caption{{\it Top}: the differential radio luminosity function in a $P \times
\phi(P)$ form for BL Lacs (black) in two redshift bins ($0 - 0.26$; dashed
and $0.26 - 0.73$; dotted lines) and FSRQ in three redshift bins: $0.2 -
0.65$ (red), $0.65 - 0.90$ (green), $0.90 - 1.22$ (blue) and at $z = 0$
(magenta).  The latter is based on the $z \le 1$ sub-sample and was
de-evolved to zero redshift using the appropriate evolutionary parameter
(see \S~\ref{fslf} and Tab. \ref{tab2}). Error bars correspond to $1\sigma$
Poisson errors \citep{geh86}. {\it Bottom}: the integral radio luminosity
function of BL Lacs in two redshift bins (dashed and dotted lines) and FSRQ
in three redshift bins and at $z = 0$.\label{blfs}}
\end{figure}

DXRBS allows us to compare the LFs of BL Lacs and FSRQ
within the same sample, which has obvious advantages.
Given the redshift dependence of the FSRQ LF (Fig. \ref{lffs}) and the fact
that BL Lacs reach only $z \sim 1$, we do this in Fig. \ref{blfs} for the
BL Lac sample, split into two equally large sub-samples at $z =
0.26$\footnote{We have assumed that the sources without redshift are
equally split between the two sub-samples and re-normalized the LFs
accordingly.}, and the $z \le 1$ FSRQ. In the latter case, we show the
three lowest redshift bins of Fig. \ref{lffs} and the LF of the $z \le 1$
sources de-evolved to zero redshift using the appropriate evolutionary
parameter (see Tab. \ref{tab2}).

%fig. 14

A few interesting points can be made: 1. the different evolutionary
properties of FSRQ and BL Lacs are visually apparent. Namely, splitting the
BL Lac LF into two redshift bins is equivalent to a simple luminosity
split, with the two LFs overlapping without discontinuity 
\citep[see][for a similar result in the X-ray band]{bec03}. The FSRQ
LFs, on the other hand, clearly display an ``evolution'', with the LFs in
different redshift bins being shifted with respect to one another, as
already discussed in \S~\ref{fslf}; 2. BL Lacs are $\sim 50$ times more
numerous than FSRQ, the former having a total number density $\sim 1,100$
Gpc$^{-3}$ for $P_{\rm r} > 7 \times 10^{23}$ W/Hz, the latter having a
total number density $\sim 23$ Gpc$^{-3}$ for $P_{\rm r} > 4 \times
10^{24}$ W/Hz. This is not simply due to the fact that BL Lacs reach lower
powers, as the BL Lac number density above $4 \times 10^{24}$ W/Hz is $\sim
240$ Gpc$^{-3}$, i.e., a factor $\sim 10$ larger than that of FSRQ above
the same radio power, due to the fact that the FSRQ LF is flatter than that of
BL Lacs ($\phi(P_{\rm r}) \propto P_{\rm r}^{-1.7}$ vs. $\phi(P_{\rm r})
\propto P_{\rm r}^{-2.1}$). The fact that FSRQ are $\approx 8$ times {\it
more} abundant in our sample (\S~\ref{counts}) is due to the fact that FSRQ
evolve, which means they are more luminous and their fluxes are
``boosted'', and to our flux limit. BL Lacs are in fact expected to
``catch up'' at lower radio fluxes, becoming the dominant blazar class
below $\approx 2$ mJy \citep{pau92}; 3. all of the above, including the
rough overlap between FSRQ and BL Lac LF in the $10^{26} \la P_{\rm r} \la
10^{27}$ W/Hz regime, is in perfect accordance with the unified schemes
of \cite{up95}
\citep[see also][]{pad92}. Namely, a scenario in which BL Lacs are beamed
FR Is and FSRQ are beamed FR IIs explains the larger number densities,
lower radio powers, and steeper LF of the former (see also
Figs. \ref{lfblbeaming} and \ref{lffsbeaming}).

Finally, the observed LFs constrain a possible evolutionary link between
the two classes.  It has in fact been suggested \citep[e.g.,][]{cav02,bo02}
that a ``genetic'' link might be present between FSRQ and BL Lacs, with
some of the former switching from a short-lived, high accretion rate regime
to a long-lived, low accretion rate regime. This would imply a decrease in
the FSRQ number density at lower redshifts, which is not seen and goes also
against the evidence discussed in \S~\ref{evolution} (see Fig.
\ref{vevabanded}).  Fig. \ref{blfs} shows in fact that the FSRQ LF evolves
smoothly to $z = 0$, while that of BL Lacs has no redshift dependence. One
could then infer that any evolutionary connection between FSRQ and BL Lacs
has then to be limited and cannot affect the bulk of the blazar
population. One caveat, however, is that radio power makes up a very small
fraction of the total, bolometric luminosity, and is also not likely to be
related to the episodes of large accretion of cold gas discussed by
\cite{cav02} and \cite{bo02}.

\section{Summary and Conclusions}\label{summary}

We have used a well-defined, complete sample selected from the Deep X-ray
Radio Blazar Survey (DXRBS) to probe the radio number counts, evolution,
and luminosity functions of blazars down to 5 GHz fluxes ($\sim 50$ mJy)
and powers ($\sim 10^{24}$ W/Hz) about one order of magnitude deeper than
previously available. Our sample includes 129 flat-spectrum radio quasars
(FSRQ) and 24 BL Lacs detected in the radio and X-ray bands over $\la
2,000$ deg$^2$. Great care has been taken in assessing the completeness of
the sample and the effects of its double X-ray/radio selection, which are
relevant for other on-going surveys as well. Our main results can be
summarized as follows:

\begin{enumerate} 
\item Our number counts agree with previous estimates at higher radio
fluxes; the surface densities of BL Lacs and FSRQ reach $\sim 0.06$
deg$^{-2}$ and $\sim 0.6$ deg$^{-2}$ respectively at $f_{\rm 5GHz} \sim 50$
mJy.
\item The two blazar sub-classes have different evolutionary
properties. FSRQ evolve as strongly as optically selected quasars, at least
up to redshift $\approx 1.5$, with evidence of no evolution and a 
decline in space density at higher redshifts for high-power sources. 
BL Lacs, on the other hand, do not
evolve. This is true also for high-energy peaked BL Lacs (HBL), at variance
with some previous results based on X-ray selected samples, which had found
evidence of negative evolution. FSRQ and BL Lac redshifts are also widely
different, with the former covering the $0.2 - 4.7$ range with $\langle z
\rangle = 1.82\pm0.08$ and the latter having $0.04 \le z \le 0.73$ and
$\langle z \rangle = 0.26\pm0.04$. 
\item The observed radio luminosity functions are in good agreement with
the predictions of unified schemes, with FSRQ getting close to their
expected minimum power ($\approx 5 \times 10^{24}/(H_0/50)^2$ W/Hz). The
idea that blazars are radio galaxies seen with their jets close to our line
of sight seems then to work also at relatively low powers.
\item Despite the fact that the large majority of DXRBS blazars are FSRQ, 
BL Lacs are intrinsically $\sim 50$ times more numerous, again
in agreement with unified schemes.
\item The observed relative numbers of HBL and LBL are different
from those predicted by the so-called "blazar sequence". The LBL/HBL ratio
is roughly constant and $\sim 6$ in the radio band, instead of the
predicted decrease down to $\sim 2$ at our flux limits. The opposite
behavior is seen in the X-ray band, where a marked (sevenfold) {\it
increase} in the LBL/HBL ratio is observed going to lower fluxes, instead
of the slight {\it decrease} expected. The available evidence supports a
scenario in which HBL are intrinsically a small minority ($\approx 10\%$)
of all BL Lacs.
\end{enumerate}  

\acknowledgments

This paper is based on observations collected at the European Southern
Observatory, Chile (ESO proposals N. 58.B-0481, 59.B-0289, 60.B-0313,
61.B-0288, 62.P-0257, 63.P-0535, 65.P-0130, 66.B-0217, 67.B-0222,
71.A-0304), Kitt Peak National Observatory, and the Australia Telescope
Compact Array. EP acknowledges support from NASA grants NAG5-9995 and
NAG5-10109 (ADP) and NAG5-9997 (LTSA). H.L. acknowledges financial support
from the Deutsche Akademie der Naturforscher Leopoldina grant BMBF-LPD
9901/8-99. We acknowledge Matteo Perri for his help in the evaluation of
the WGACAT sky coverage and Piero Rosati for useful discussions. This
research has made use of the NASA/IPAC Extragalactic Database (NED), which
is operated by the Jet Propulsion Laboratory, California Institute of
Technology, under contract with the National Aeronautics and Space
Administration.

{\it Facilities:} \facility{ATCA}, \facility{Parkes}, \facility{GBO:42.6m},
\facility{VLA}, \facility{ESO:3.6m (EFOSC2)}, \facility{VLT:Kueyen
(FORS1)}, \facility{VLT:Antu (FORS2)}, \facility{Max Plank:2.2m (EFOSC1)},
\facility{KPNO:2.1m}, \facility{ROSAT}

\end{document}